\documentclass[epj]{svjour}
\usepackage{latexsym}
\usepackage{graphics}
\usepackage{amsmath}
\usepackage{flushend}
\usepackage{cuted}
\usepackage{subfig}
\usepackage{comment}
\usepackage{gensymb}

\begin{document}
\title{$\eta$ and $\omega$ mesons as new degrees of freedom in the intranuclear cascade model INCL}
\author{J.-C. David\inst{1} \and A. Boudard\inst{1} \and J. Cugnon\inst{2} \and J. Hirtz\inst{1,3} \and S. Leray\inst{1} \and D. Mancusi\inst{4} \and J. L. Rodriguez-Sanchez\inst{1}
}                     
%
%
\institute{Irfu, CEA, Universit\'e Paris-Saclay, 91191 Gif-sur-Yvette, France \and University of Li\`ege, AGO Department, all\'ee du 6 ao\^ut 17, B\^at. B5, B-4000 Li\`ege 1, Belgium \and Center for Space and Habitability, Universit\"{a}t Bern, CH-30 12 Bern, Switzerland \and Den-Service d'\' etude des r\' eacteurs et de math\'ematiques appliqu\' ees (SERMA), CEA, Universit\'e Paris-Saclay, 91191 Gif-sur-Yvette, France}
%
%
\abstract{
The intranuclear cascade model INCL (Li\`ege Intranuclear Cascade) is now able to simulate spallation reactions induced by projectiles with energies up to roughly 15 GeV. This was made possible thanks to the implementation of multipion emission in the NN, $\Delta$N and $\pi$N interactions. The results obtained with reactions on nuclei induced by nucleons or pions gave confidence in the model. A next step will be the addition of the strange particles, $\Lambda$, $\Sigma$ and Kaons, in order to not only refine the high-energy modeling, but also to extend the capabilities of INCL, as studying hypernucleus physics. Between those two versions of the code, the possibility to treat the $\eta$ and $\omega$ mesons in INCL has been performed and this is the topic of this paper. Production yields of these mesons increase with energy and it is interesting to test their roles at higher energies. More specifically, studies of $\eta$ rare decays benefit from accurate simulations of its production. These are the two reasons for their implementation. Ingredients of the model, like elementary reaction cross sections, are discussed and comparisons with experimental data are carried out to test the reliability of those particle productions. 
%
} 
\maketitle
\section{Introduction}
\label{Intro}

Nuclear reactions between a light particle ({\it e.g.} hadron) and a nucleus have been extensively studied. In the last twenty years, for incident energies from $\sim$100 MeV up to a few GeV, great improvements were obtained, as shown by the two international benchmarks carried out in mid-nineties \cite{BLA93,MIC97} and 2010 \cite{LER11}. The studies of those reactions, which take place in space due to the cosmic rays as well as in accelerators, has been triggered mostly by transmutation of nuclear wastes. This explains the energy domain and the focus on residual nucleus and neutron production, even if light charged particles (proton, deuteron, ..., and also pion) were also studied. Some codes can already simulate those types of reactions for higher incident energies, taking into account other particles than nucleons and pions, since new particles appear when energy increases. Around say 10 GeV two groups of models can be used to reproduce these reactions. A first group includes the high energy models, often based on a String model ({\it e.g.} \cite{NIL87,SJO06}, from the TeV (or higher) down to a few GeV, and a second group with the BUU ({\it e.g.} \cite{AIC85,LI89,BUS12}), QMD ({\it e.g.} \cite{AIC91,NII95,BAS98,HAR98}) and intranuclear cascade (INC) ({\it e.g.} see Table 1 in \cite{DAV15}) models where the extension to higher energies needs new ingredients.

This article is about the extension of the intranuclear cascade model INCL toward the high energies (10-15 GeV). In 2011 S. Pedoux and J. Cugnon \cite{PED11} did the main part of the work by implementing the multiple pion production processes in the elementary interactions (NN, $\Delta$N and $\pi$N). The idea rested on the two facts that i) when the energy goes up new particles are produced, {\it i.e.} especially new resonances, which will decay mainly in pions and nucleon in a time much shorter than the duration of the cascade, and ii) information on those resonances (masses, widths, related cross sections) are not always well known and moreover overlaps exist between resonances making the choice awkward. This multipion production model in INCL~\cite{BOU13,MAN14} leads to good results regarding pion production \cite{MAN17} compared to experimental data and to other models. 

Nevertheless, even though pions are the main particles produced, some others like $\eta$ and $\omega$ mesons and the strange particles (kaons and hyperons) can appear as well. Implementation of these particles should not significantly change the global features of the reactions (residual nucleus, neutron and light charged particle production), but first this has to be confirmed, and second those particles can be new fields of study. Considering strange particles, hypernuclei can be then studied. While the implementation of kaons and hyperons is in progress in INCL and will be soon addressed in another paper, this one is dedicated to the $\eta$ and $\omega$ mesons. The motivations, besides the completeness of the code, were the quantification of the role of $\eta$ and $\omega$ on the pion production, the fact that they are sources of dileptons, which are probes of the nuclear matter, the need to get a good knowledge of $\eta$ production to be able to study rare decays indicating violation of some special conservation laws (\cite{KUP11,GAT16}), and finally a necessary step toward the strange sector implementation.

The needed ingredients for the INCL code are discussed in sect. \ref{Inputs}. The main ones are the elementary cross sections related to processes where those new mesons are involved (sects. \ref{XS} and \ref{Output}), but decays and in-medium potentials are also addressed (sects. \ref{Decays} and \ref{Potentials}). Section \ref{Results} is devoted to the results obtained and compared to experimental data, with discussions and analyses related to the input ingredients. Conclusions are given in sect.~\ref{Concl}.

\section{Elementary ingredients}
\label{Inputs}

Necessary ingredients, when adding new particles in an INC code, in addition to the own properties (mass, charge), are elementary cross sections, decays and in-medium potentials. Reaction cross sections are used to simulate processes where the particle plays a role (production, absorption and scattering), and differential cross sections to characterize the output channels of those processes. The lifetime of the $\eta$ ($\sim$ 150000 fm/c) is much larger than the duration of the intranuclear cascade (around 70 fm/c), thus its decay is considered only at the end of the cascade, if it has been emitted from the nucleus. This is not the case of the $\omega$ (lifetime $\sim$ 23~fm/c), whose decay must be taken into account also during the cascade. Regarding the potential felt by particles in the nuclear medium, and relevant mainly at low energy, unfortunately few information are available. Those input ingredients are based on experimental data, and, when they are missing, on symmetries ({\it e.g.} isospin), models, hypotheses or extrapolations. The following sections describe in detail each topic.

\subsection{Cross sections}
\label{XS}

Elementary reaction cross sections considered are those that characterize processes in which an $\eta$ or an $\omega$ is involved. Then production of those particles is governed by the NN and $\pi$N interactions. Although there is much less $\pi$ than nucleons, their related cross sections are almost one order of magnitude higher than those of nucleons. Once produced mesons can undergo elastic scattering, which is considered so as to better reproduce the multiple scattering, as well as absorption. Only absorption on one nucleon is accounted for.


\subsubsection{Production}
\label{XS-prod}

Particle production can be exclusive or inclusive, since, when energy increases, additional particles may be produced. This has been taken into account in NN reactions, but not in $\pi$N. 
\medskip

\underline{$\pi$N $\rightarrow$ $\eta$($\omega$)N}
\medskip

\begin{figure}
\centering
\resizebox{0.4\textwidth}{!}{\includegraphics{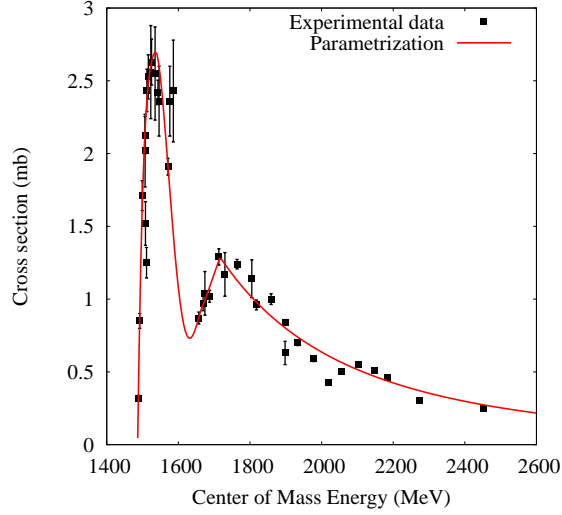}}
\caption{$\pi^-p \rightarrow \eta n$ reaction cross section. Our parametrization is given by the red solid line and the experimental data  are those of Richards {\it et al.} \cite{RIC70}, Prakhov {\it et al.} \cite{PRA05}, Brown {\it et al.} \cite{BRO79} and Deinet {\it et al.} \cite{DEI69}.}
\label{Fig1}
\end{figure}

The parametrization of the $\eta$ production cross section, via the $\pi$N reaction, is based on a fit of experimental data. The data used (\cite{RIC70, PRA05, BRO79, DEI69}) are those studied in the paper of J. Durand {\it et al.} \cite{DUR08}, where their reliability was investigated. Equation \ref{eq:piN2eta} and fig. \ref{Fig1} give the results of the fit performed for $\sigma(\pi^-p \rightarrow \eta n)$ where the energy range was divided in several domains. The high energy domain ($E_{cm} \geq 1714$~MeV)) is function of the laboratory momentum, unlike the others which are function of  the center of mass energy. Actually the parametrization of the high energy domain is the one of Cugnon {\it et al.} \cite{CUG90}, while the formula of the lower parts were defined to match at best the experimental data, especially around the resonance N(1535). The other $\sigma(\pi N \rightarrow \eta N)$ are derived from isospin symmetry.
\begin{equation}\label{eq:piN2eta}
\sigma(\pi^-p \rightarrow \eta n) =
\begin{cases}
\sum_{i=0}^4a_iE_{cm}^i & threshold \leq E_{cm} < 1714.0  \\ 
1.47\ P_{Lab}^{-1.68}      & 1714.0 \leq E_{cm}  \\
\end{cases}
\end{equation}

($a_i$ parameters are given in appendix~\ref{Apxspar1})

\fbox{$E_{cm}:$ MeV; $P_{Lab}:$ MeV/c; $\sigma: mb$} 

\medskip

\begin{figure}
\centering
\resizebox{0.4\textwidth}{!}{\includegraphics{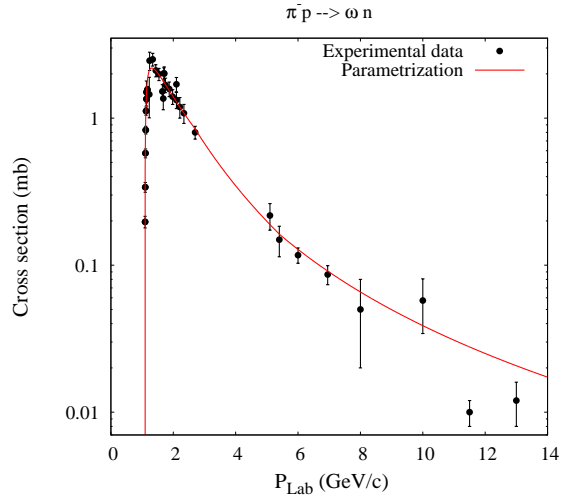}}
\caption{$\pi^-p \rightarrow \omega N$ reaction cross section. Our parametrization is given by the red solid line and the experimental data are those from Landolt-B\"{o}rnstein \cite{BAL88} (black points).}
\label{Fig2}
\end{figure}

The parametrization used for $\sigma(\pi N \rightarrow \omega N)$ is an improved version of the one of Cugnon {\it et al.} \cite{CUG90} where a parameter was slightly modified (1.095 GeV changed by 1.0903 GeV) to better account for the threshold of the reaction (eq. \ref{eq:piN2omega}). This is important for the reverse reaction, absorption, which is obtained with the detailed balance (see sect. \ref{XS-abs}). Figure \ref{Fig2} shows the result for $\sigma(\pi^-p \rightarrow \omega n)$.

\begin{eqnarray}\label{eq:piN2omega}
\sigma(\pi^-p \rightarrow \omega n) = 13.76 \hspace{0.1 cm} \frac{(P_{Lab}-1.0903)}{(P_{Lab}^{3.33} - 1.07)}  & 1.0903 \leq P_{Lab} 
\end{eqnarray}

\fbox{$P_{Lab}:$ GeV/c; $\sigma:$ mb} 
\bigskip

\underline{NN $\rightarrow$ $\eta$($\omega$)+X}
\medskip

The situation is different with the NN reactions. The inclusive channel becomes quickly important and unfortunately experimental data are very rare. The case of $\eta$ is discussed first and then the $\omega$. 

Considering the exclusive production, NN $\rightarrow$ NN$\eta$, experimental data for the pp channel allow a more or less reliable parametrization up to $E_{cm}\sim 5$ GeV. For the pn $\rightarrow$ pn$\eta$ reaction the data available are only known close to the threshold and a scale factor, compared to the pp channel, of $\sim 6.5$ appears \cite{CAL98}. This factor seems to exhibit the predominant role of an isovector exchange ($\pi,\rho$), since in this case a factor 5 is theoretically expected \cite{HAN04}. However, we performed our own parametrization. We aimed at matching the experiments at threshold and at considering that $\eta$ production in the np reaction should be equivalent to the production in the pp reaction beyond the resonance region (we decided beyond 3.9~GeV). In-between we assumed that the shapes should be also not too different. The lack of experimental data and models for $np~\rightarrow~np\eta$ beyond the threshold drive us to those kinds of hypotheses. This has to be taken into consideration when analysing calculation results, since it could lead to significant uncertainties. 
Moreover, at threshold, the reaction pn $\rightarrow$ d$\eta$ is dominant and thus must be taken into account. Since deuteron is not a degree of freedom in INCL, this cross section is just added to the {\it pure} pn channel (d = np). Equations \ref{eq:pp2ppeta}, \ref{eq:np2npeta}, and \ref{eq:pn2deta} give the parametrizations ($a_i$ and $b_i$ parameters are given in appendices~\ref{Apxspar2} and \ref{Apxspar3}). 
\begin{equation}\label{eq:pp2ppeta}
\sigma(pp \rightarrow pp\eta) =  \sum_{i=0}^5a_iE_{cm}^i  \hspace{1 cm} threshold \leq E_{cm}  \\
\end{equation}
\begin{equation}\label{eq:np2npeta}
\sigma(np \rightarrow np\eta) = 
\begin{cases}
\sum_{i=0}^4b_iE_{cm}^i          & threshold \leq E_{cm}  < 3.9 \\ 
\sigma(pp \rightarrow pp\eta) &  3.9 \leq E_{cm} \\ 
\end{cases}
\end{equation}

\fbox{$E_{cm}:$ GeV; $\sigma: \mu$b}

\begin{equation}\label{eq:pn2deta}
\sigma(pn \rightarrow d\eta) = -1.02209 \hspace{0.07 cm} 10^{4}\hspace{0.07 cm} E_{cm}^2 + 5.12273 \hspace{0.07 cm} 10^{4}\hspace{0.07 cm} E_{cm} - 6.40980 \hspace{0.07 cm} 10^{4}\hspace{0.07 cm}   
\end{equation}
\fbox{$E_{cm}:$ GeV; $\sigma: \mu$ b (when $\sigma < 0$ ($E_{cm}$ $>$ 2.6 GeV), $\sigma$ set to 0)}
\bigskip

\begin{figure}
\centering
\resizebox{0.45\textwidth}{!}{\subfloat{\label{Fig3a}\includegraphics{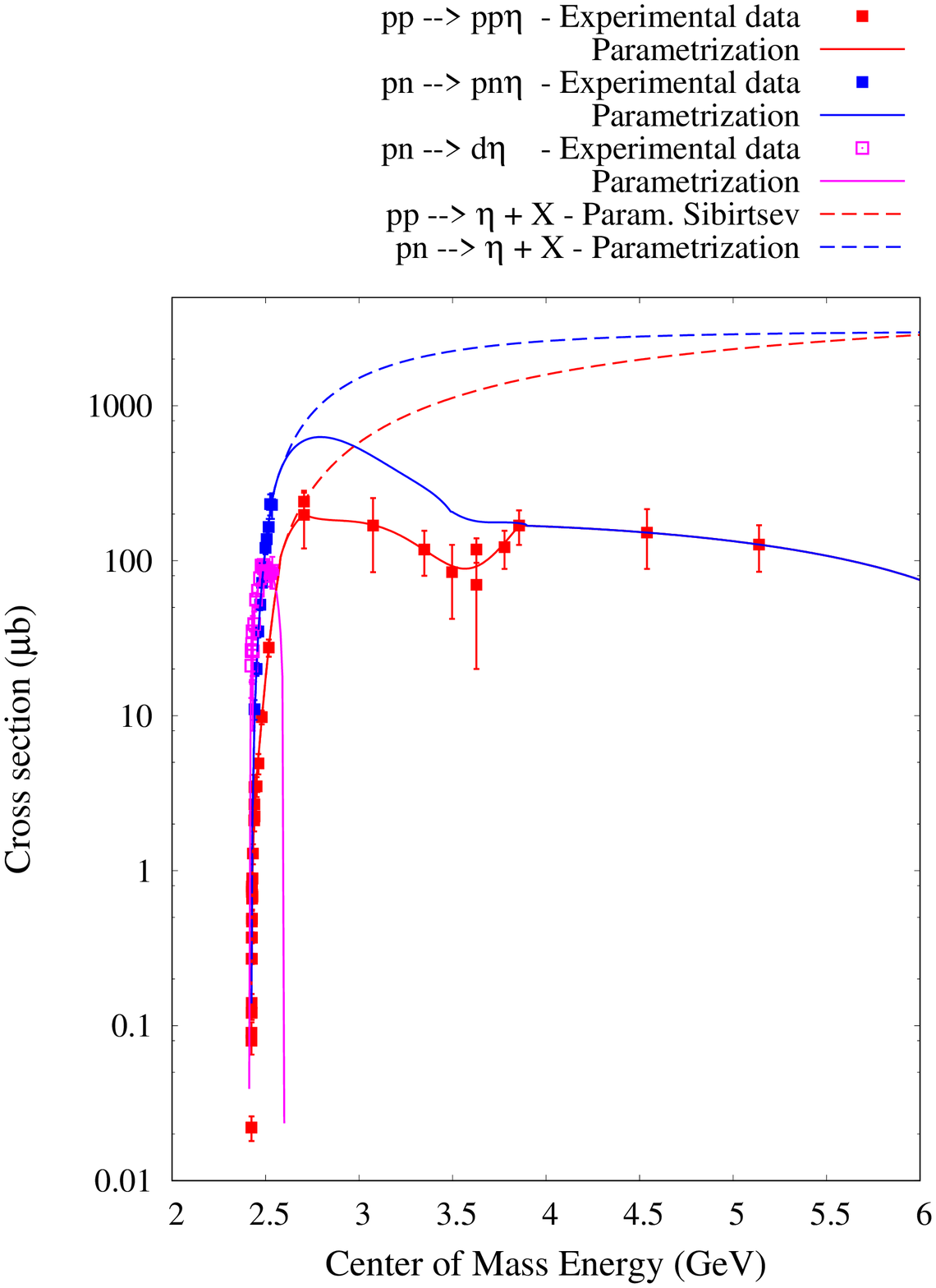}}}
\hspace{1 cm}
\resizebox{0.25\textwidth}{!}{\subfloat{\label{Fig3b}\includegraphics{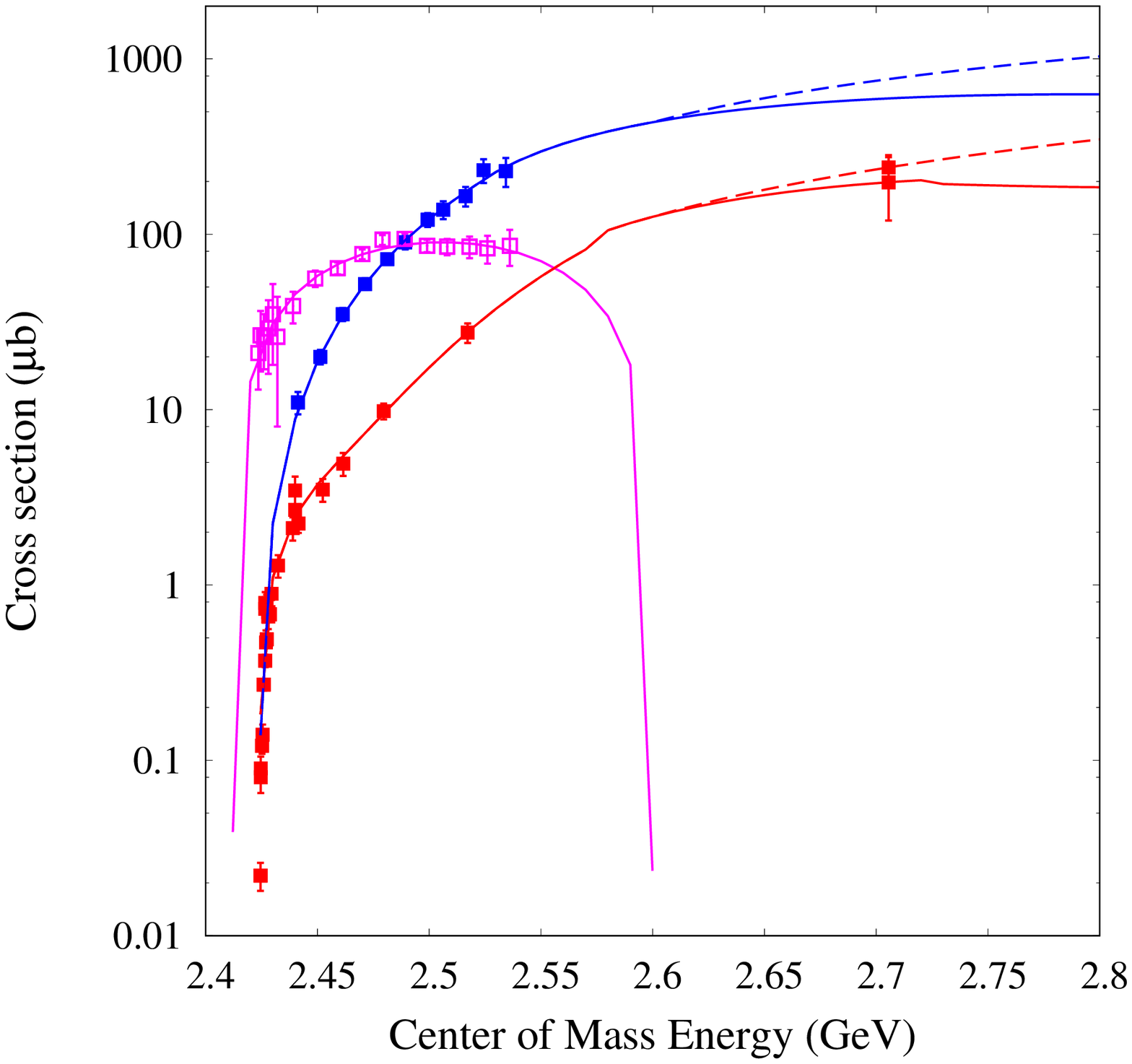}}}
\caption{$NN \rightarrow \eta + X$ reaction cross sections. Exclusive (solid lines) and inclusive (dashed lines) production are plotted. Our parametrizations for the exclusive cases are fitted on experimental data and the ones for the inclusive cases are those of Sibirtsev \cite{SIB97}. Experimental data  come from \cite{SMY00,CAL96,BER93,HIB98,CHI94,BAL88} for the pp channel (red), \cite{CAL98} for the {\it pure} pn channel (blue), and \cite{CAL97,CAL98b} for the deuteron channel (magenta). The figure on the right side is a focus on the low-energy part.}
\label{Fig3}
\end{figure}

The inclusive reactions were studied elsewhere already, in particular in Sibirtsev {\it et al.} \cite{SIB97} where a parametrization can be found for the pp channel. Nevertheless, since this inclusive production parametrization differs from the exclusive one at threshold (approximatively up to 2.6 GeV in center of mass energy), we decided to use the Sibirtsev formula only beyond a given energy (chosen at 3.05 GeV) and to connect smoothly the two energy domains. The case of the np channel is once again a problem: there is no experimental data, nor available model. Our two rules here were first an inclusive cross section different of the exclusive one only above 2.6 GeV (like in pp) and second a similar cross section to pp beyond 6.25 GeV. Using two different values in the exclusive (3.9 GeV) and inclusive (6.25 GeV) cases from where the pp and np channels are supposed to give the same cross section is not totally satisfactory, but corresponds to reasonable compromise. The factor 6.5 observed from threshold up to almost 2.6~GeV seems hard to reduce to 1 at 3.9 GeV (as in exclusive case). These questionable points are discussed further in sect.~\ref{Results}.  Formulas are given below (eqs. \ref{eq:pp2eta} and \ref{eq:np2eta}) and the cross sections (exclusive and inclusive) are plotted in fig.~\ref{Fig3}.

Now remains the question: what does X mean in the inclusive reactions? 

\begin{figure}
\centering
\resizebox{0.5\textwidth}{!}{\includegraphics{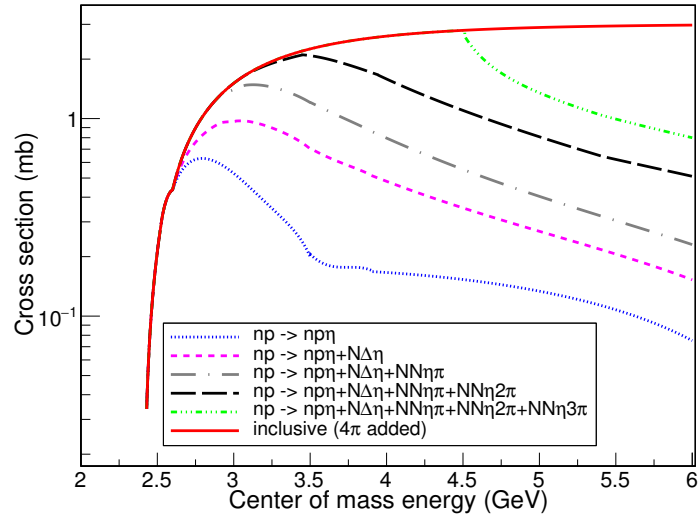}}
\caption{$pn \rightarrow NN\eta + x\pi$ reaction cross sections. The total inclusive cross section is given by the red solid line, the exclusive production by the blue line, the addition of the $\Delta$ contribution by the magenta line, the addition of the one pion contribution by the brown line, the addition of the two pions contribution by the black line, and the addition of the three pions contribution by the green line. The gap between the green line and the red solid line represents the four pions contribution.}
\label{Fig4}
\end{figure}

No experiment gives information on the content of X. Keeping in mind that $\eta$ (and $\omega$) production is much less important than pion production, and that, whatever the particle (resonance) created, most of the decay products are pions and nucleons, the X has been supposed to be "NN + x$\pi$" in our study . In addition, this solution was straightforward to implement, because it is nothing but the multiple pion production mechanism already put in INCL \cite{PED11}. This was of course possible because the $\eta$ (and $\omega$) isospin is 0 and keeps unchanged all equations based on the isospin symmetry concerning the multipion channels developed in \cite{PED11}. Only the threshold of the various pion emission has been moved, due to the needed minimal energy for $\eta$ production. The value, 581.437 MeV, has been determined from the comparison of our exclusive cross sections and the inclusive cross sections from \cite{SIB97}, and it corresponds to the center of mass energy, 2.6~GeV, where the two parametrizations separate from each other. Figure~\ref{Fig4} shows the case of pn reactions with all the channels. 

\begin{equation}\label{eq:pp2eta}
\sigma(pp \rightarrow \eta + X) = 
\begin{cases}
\sigma(pp \rightarrow pp\eta) &  E_{cm}  \leq 2.6
\\
- 3.2729 \hspace{0.07 cm} 10^{2}\hspace{0.07 cm}E_{cm}^3 +2.87\hspace{0.07 cm} 10^{3}\hspace{0.07 cm}E_{cm}^2 - 7.2293\hspace{0.07 cm} 10^{3}\hspace{0.07 cm}E_{cm} + 5.2733\hspace{0.07 cm} 10^{3} &  2.6 \leq E_{cm}  < 3.05
\\
2.5\hspace{0.07 cm} 10^{3}*((E_{cm}^2/5.88)-1)^{1.47} * (E_{cm}^2/5.88)^{-1.25} &  3.05 \leq E_{cm}
\end{cases}
\end{equation}

\begin{equation}\label{eq:np2eta}
\sigma(np \rightarrow \eta + X) = 
\begin{cases}
\sigma(np \rightarrow np\eta) &  E_{cm}  \leq 2.6
\\
\sigma(pp \rightarrow \eta + X) * e^{-(5.5315/E_{cm} + 0.8850)} &  2.6 \leq E_{cm}  < 6.25
\\
2.5\hspace{0.07 cm} 10^{3}*((E_{cm}^2/5.88)-1)^{1.47} * (E_{cm}^2/5.88)^{-1.25} &  6.25 \geq E_{cm}
\end{cases}
\end{equation}
\fbox{$E_{cm}:$ GeV; $\sigma$: $\mu$b}

\begin{figure}
\centering
\resizebox{0.45\textwidth}{!}{\includegraphics{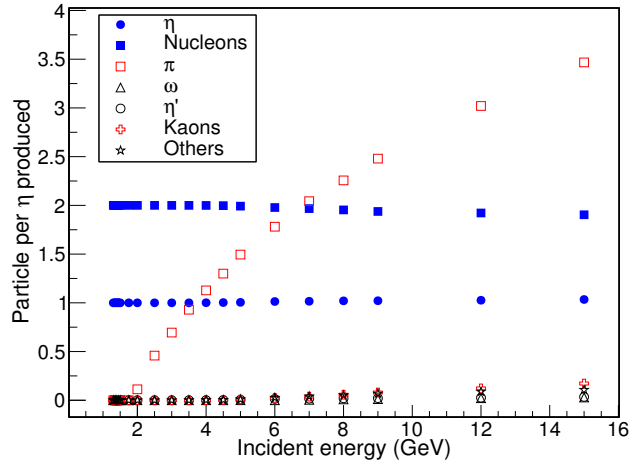}}
\caption{Multiplicities of particles associated with the emission of one $\eta$ in the NN reactions, given by the Fritiof model in Geant4.}
\label{Fig5}
\end{figure}
\begin{figure}
\centering
\resizebox{0.45\textwidth}{!}{\includegraphics{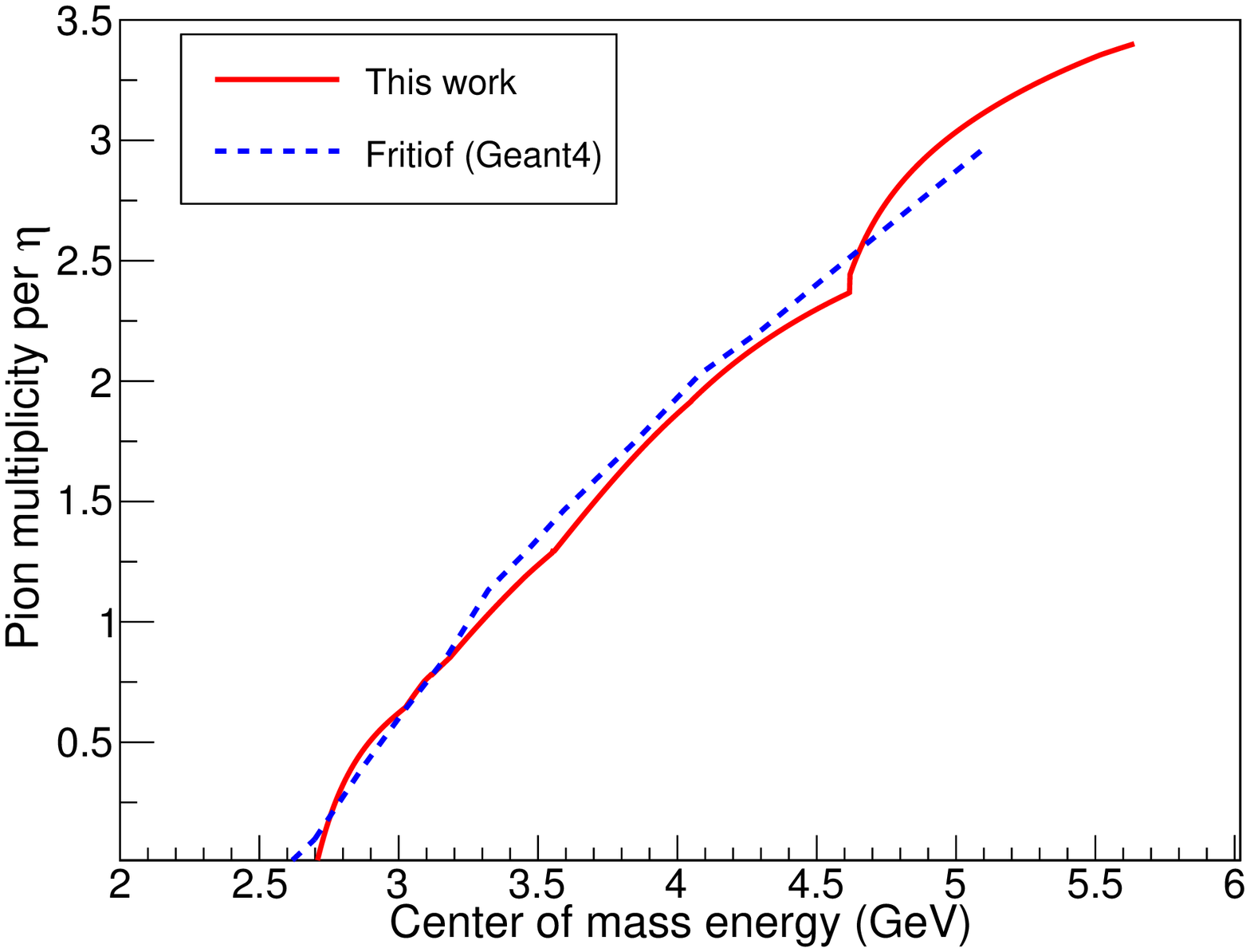}}
\caption{Pion multiplicities versus center of mass energy for the reaction $pn \rightarrow NN\eta + x\pi$. Our parametrization is the solid red line and the results of Fritiof the dashed blue line.}
\label{Fig6}
\end{figure}

As said just previously the multiple pion mechanism is the one already applied in INCL and the one-pion emission in the isospin T=1 channel is governed by the $\Delta$ resonance. This is possibly no more the case considering the energy when $\eta$ (or $\omega$) are produced, and the pion could be directly produced. Since the main goal is to give to the $\eta$ a realistic energy, we assume that  the number of associated pion is important, not the way to get them. Then, for sake of simplicity, the $NN\rightarrow N\Delta\eta(\omega)$ has been kept. Moreover, fig.~\ref{Fig5} shows the particles, type and number, created when one $\eta$ is produced by the Fritiof model used in GEANT4 and the comparison between this Fritiof model and our results for the pion multiplicity associated with emission of an $\eta$ shown in fig. \ref{Fig6}. Both results give confidence in the meaning of X, {\it i.e.} up to $\sim$ 15 GeV the final products associated to one $\eta$ are two nucleons and pions, whose multiplicity of the latter grows with energy and is well reproduced by the multipion emission process. The curious shape, for INCL, around $E_{cm} =$ 4.5~GeV is only a consequence of the multiple pion cross section parametrization, and more precisely of the way the $4\pi$ emission channel starts at $E_{cm} \approx$ 4.5~GeV, with a quick increase (fig.~\ref{Fig4}). 

\begin{figure}
\centering
\resizebox{0.45\textwidth}{!}{\includegraphics{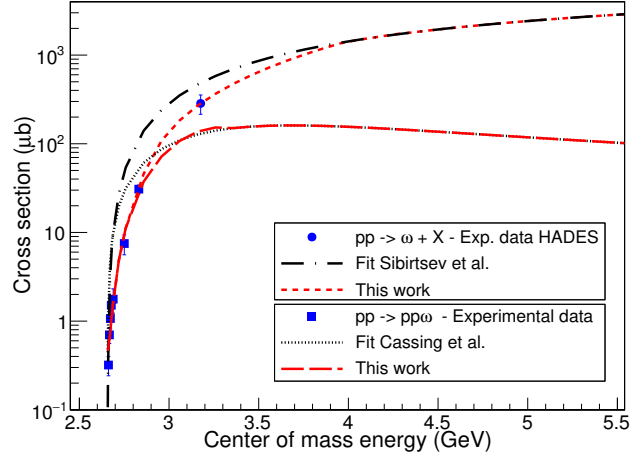}}
\caption{$pp \rightarrow \omega + X$ reaction cross sections. Exclusive and inclusive production are plotted. Our parametrization for the exclusive case is the dashed red line and for the inclusive case the dotted red line. Also plotted are the parametrizations of Cassing \cite{SIB96} (exclusive) and Sibirtsev \cite{SIB97} (inclusive). Experimental data come from \cite{HIB99,ABD01,BAL88} for the exclusive channel and \cite{RUS10} for the inclusive one. More details in text.}
\label{Fig7}
\end{figure}

A similar procedure has been applied to $\omega$ production. Equations \ref{eq:pp2omega} and \ref{eq:pp2omegaX} give parametrizations for the exclusive and inclusive processes in  the pp channel. The formula given by Cassing \cite{SIB96} has been used for the exclusive case, except at threshold (below $E_{cm} = 3.0744$ GeV) where a new fit has been produced to match better the experimental data. Regarding the inclusive process the formula of Sibirtsev~\cite{SIB97} is applied above $E_{cm} = 4$ GeV and between 2.802~GeV and 4 GeV a new parametrization has been used to take into account one experimental point from the HADES collaboration \cite{RUS10}. Below $E_{cm} = 2.802$~GeV only the exclusive process occurs. For the np channel a factor 3, from the pp channel, has been applied (compared to the factor 6.5 in the $\eta$ case). This factor is a bit arbitrary and relies on the experimental value of $\sim$3 obtained in \cite{BAR04}. Figure \ref{Fig7} compares the parametrizations to experimental data. Less attention has been paid to the $\omega$ meson, since unfortunately no experimental data exist on the production from a nucleus, as far as we know, and so no possibility to test INCL.

\begin{equation}\label{eq:pp2omega}
\sigma(pp \rightarrow pp\omega) =  
\begin{cases}
 -1.2081\hspace{0.07 cm} 10^{3} \hspace{0.07 cm}E_{cm}^3 + 1.0773\hspace{0.07 cm} 10^{4}\hspace{0.07 cm}E_{cm}^2  - 3.1661\hspace{0.07 cm} 10^{4}\hspace{0.07 cm}E_{cm} + 3.0729\hspace{0.07 cm} 10^{4} & threshold\leq E_{cm}  < 3.0744 \\
330.\hspace{0.07 cm} \frac{E_{cm}-\sqrt{7.06}}{1.05+(E_{cm}-\sqrt{7.06})^2} & E_{cm}  \geq 3.0744 \\
\end{cases}
\end{equation}
\begin{equation}\label{eq:pp2omegaX}
\sigma(pp \rightarrow \omega + X) =  
\begin{cases}
\sigma(pp \rightarrow pp\omega) &  E_{cm}  < 2.802 \\
  568.5254 \hspace{0.07 cm}E_{cm}^2 - 2694.0450 \hspace{0.07 cm}E_{cm} + 3106.2470 &  2.802\leq E_{cm}  < 4.  \\
2500. \hspace{0.07 cm}\frac{(E_{cm}^2/7.06-1)^{1.47}}{(E_{cm}^2/7.06)^{-1.11}} &  E_{cm}  \geq 4. \\
\end{cases}
\end{equation}
\fbox{$E_{cm}:$ GeV; $\sigma$: $\mu$b}

\subsubsection{Elastic scattering}
\label{XS-elas}

Elastic scattering of $\eta$ and $\omega$ on the nucleon is accounted for to treat better the energy spectrum and the final emission angle.
\medskip

\underline{$\eta$($\omega)$N $\rightarrow$ $\eta$($\omega$)N}
\medskip

In the $\eta$ case the elastic scattering parametrization is based on calculation results kindly provided by H. Kamano. He and his coworkers studied the $\pi N \rightarrow \eta N$ reactions (up to a center of mass energy of 2.1 GeV) in  the frame of a more general investigation of nucleon resonances within a dynamical coupled-channels model of $\pi N$ and $\gamma N$ reactions \cite{KAM13}. The results they obtained compared to experimental data ({\it e.g.} for $\pi N \rightarrow \eta N$ reactions) give confidence in their ANL-Osaka model and so in the extrapolation to other reactions like $\eta N\rightarrow \eta N$. Our parametrization is based on polynomials and the energy range divided in three domains (eq.~\ref{eq:etaN2etaN} where $a_i$ parameters are given in appendix~\ref{Apxspar4}). Figure~\ref{Fig8} shows the results also compared to three rare experimental data~\cite{DUD76}. The comparison is quite satisfactory. Although the extrapolation beyond 1.5~GeV/c is questionable and can add uncertainties in the angular distributions, we expect low impacts on the results, because at those energies the elastic cross sections are low and the $\eta$'s are scattered in the very forward direction (see sect.~\ref{Output-elas}).

\begin{figure}[h]
\centering
\resizebox{0.35\textwidth}{!}{\includegraphics{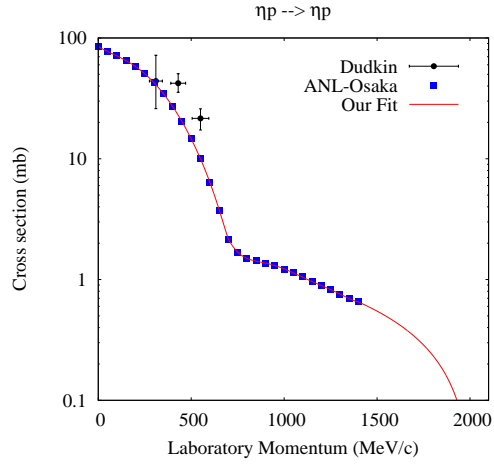}}
\caption{$\eta p \rightarrow \eta p$ reaction cross section. Our parametrization is given by the solid red line and the ANL-Osaka calculation results \cite{KAM13} are the blue squares. Experimental data from Dudkin \cite{DUD76} are plotted in black.}
\label{Fig8}
\end{figure}

\begin{equation}\label{eq:etaN2etaN}
\sigma(\eta N \rightarrow \eta N) = 
\begin{cases}
\sum_{i=1}^6a_iP_{Lab}^i  &  P_{Lab} < 2025. \\
 0                                      & 2025. \leq  P_{Lab} \\
\end{cases}
\end{equation}
\fbox{$P_{Lab}:$ MeV/c; $\sigma$: mb}
\medskip

For $\omega$ the formula of Lykasov~\cite{LYK99} has been chosen (eq.~\ref{eq:omegaN2omegaN}). Above $P_{Lab} = 10$ GeV/c this parametrization is only an extrapolation, because it is supposed to be used only in the range 10 MeV/c~-~10 GeV/c.

\begin{equation}\label{eq:omegaN2omegaN}
\sigma(\omega N \rightarrow \omega N) = 5.4 + 10.^{-0.6\hspace{0.07 cm}P_{Lab} }\\
\end{equation}
\fbox{$P_{Lab}:$ GeV/c; $\sigma$: mb}

\subsubsection{Absorption}
\label{XS-abs}

The production rate of any particle can be well simulated only if the absorption is also considered.
In the case of $\eta$ and $\omega$ meson in the nucleus, absorption on the nucleons is obviously the main channel.
\medskip

\underline{$\eta$N $\rightarrow$ $\pi$N, $\pi$$\pi$N}
\medskip

As for the elastic scattering, our parametrizations for absorption reactions are based on calculation results of the ANL-Osaka model~\cite {KAM13}. Actually, H. Kamano provided, in addition to the elastic cross section, the total cross section as well as three inelastic channels, {\it i.e.} one and two pions production ($\eta$N $\rightarrow$ $\pi$N or $\pi$$\pi$N) and Kaon-Hyperon production ($\eta$N $\rightarrow$ KY, Y=$\Lambda$ or $\Sigma$). Since the strange particles are not yet implemented in INCL, only pion production channels are taken into account. This underestimates the inelastic and total cross section, but only above 500 MeV/c first, second in a reasonable amount up to 1500 MeV/c and third only up to the strange particles will be available in INCL (next step).  Equations \ref{eq:etaN2piN} and \ref{eq:etaN2pipiN} give our parametrizations ($a_i$ and $b_i$ parameters are given in appendices~\ref{Apxspar5} and \ref{Apxspar6}) and fig. \ref{Fig9} shows comparisons between the ANL-Osaka model and our fits. It must be stressed that beyond 1~GeV ($\eta$ energy) our cross sections are only extrapolations and thus any calculation result analysis must take it into consideration.

\begin{equation}\label{eq:etaN2piN}
\sigma(\eta N \rightarrow \pi N) = 
\begin{cases}
\sum_{i=0}^6a_iP_{Lab}^i &   P_{Lab} \leq 1300. \\
\mathrm{detailed\ balance} &   1300. < P_{Lab} \\
\end{cases}
\end{equation}
\begin{equation}\label{eq:etaN2pipiN}
\sigma(\eta N \rightarrow \pi\pi N) = 
\begin{cases}
\sum_{i=0}^6b_iP_{Lab}^i  &  P_{Lab} \leq 450. \\
\sigma(450 \  MeV/c) & 450.  < P_{Lab} \leq 600. \\
\sum_{i=0}^6b_iP_{Lab}^i  &  600.  < P_{Lab} \leq 1300.  \\
\sigma(\eta N \rightarrow \pi N)   &  1300. < P_{Lab} \\
\end{cases}
\end{equation}
\fbox{$P_{Lab}: MeV/c$; $\sigma$: mb}
\medskip

\begin{figure}
\centering
\resizebox{0.45\textwidth}{!}{\includegraphics{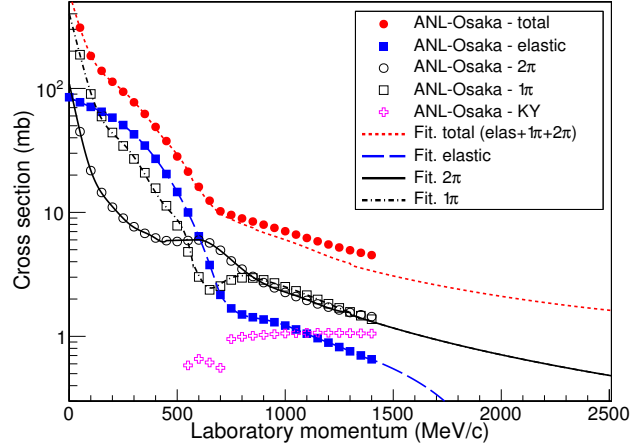}}
\caption{$\eta p \rightarrow X$ reaction cross sections. Our parametrizations are given by the lines and ANL-Osaka calculation results \cite{KAM13} by the marks.}
\label{Fig9}
\end{figure}

\underline{$\omega$N $\rightarrow$ $\pi$N, $\pi$$\pi$N}
\medskip
 
Concerning the $\omega$ meson we resorted to detailed balance for the $\omega$N~$\rightarrow$~$\pi$N reaction and obtained the $\omega$N~$\rightarrow$~$\pi\pi$N reaction by subtraction using a parametrization of Lykasov \cite{LYK99} for the inelastic reaction ($\sigma_{\omega N\ \rightarrow\ \pi\pi N} = \sigma^{inelastic}_{\omega N} - \sigma_{\omega N\ \rightarrow\ \pi N}$). Equation \ref{eq:lyk-inelastic} gives the inelastic parametrization.
\begin{equation}\label{eq:lyk-inelastic}
\sigma^{inelastic}_{\omega N} = 20. + 4./P_{Lab}\hspace{0.5cm}
\end{equation}
\fbox{$P_{Lab}:$ GeV/c; $\sigma$: mb}

\subsection{Features of the reaction products}
\label{Output}

Whatever the reaction, the final state must be characterized, {\it i.e.} type, energy and direction of the particles defined. In this topic the reactions are divided into two families according to the number of particles in the output channel: two or more.
When only two particles exist, the charges are either obvious or obtained randomly from the Clebsch-Gordan coefficients and their energies in the center of mass are given by the laws of conservation of energy and momentum. Regarding the direction, if no information is available, isotropy is assumed, while, if experimental data or calculation results (single or double differential cross sections) exist, parametrizations of the emission angles are drawn. For the cases with three or more particles, the charge of particles is obtained either with the Clebsch-Gordan coefficients or, when it is not possible, from models assuming hypotheses to remove ambiguities. Energies and directions are derived from a phase-space generator.

While a few information exist for $\eta$, almost nothing for $\omega$. 
Therefore, in the case of the $\omega$ meson, we assume isotropy for two particles, and use a phase-space generator when three or more particles are produced. Regarding the $\eta$ meson, more details are given in the following sections, according to the type of the reaction.

\subsubsection{Production}
\label{Output-prod}

The $\eta$ meson is produced through $\pi$N and NN reactions. While more than one particle can be associated to the $\eta$ with increasing energy, we consider only two particles in the final state for the $\pi$N case, for lack of information. 
\medskip

\underline{$\pi$N $\rightarrow$ $\eta$N}
\medskip

\begin{figure}
\centering
\resizebox{0.425\textwidth}{!}{\includegraphics{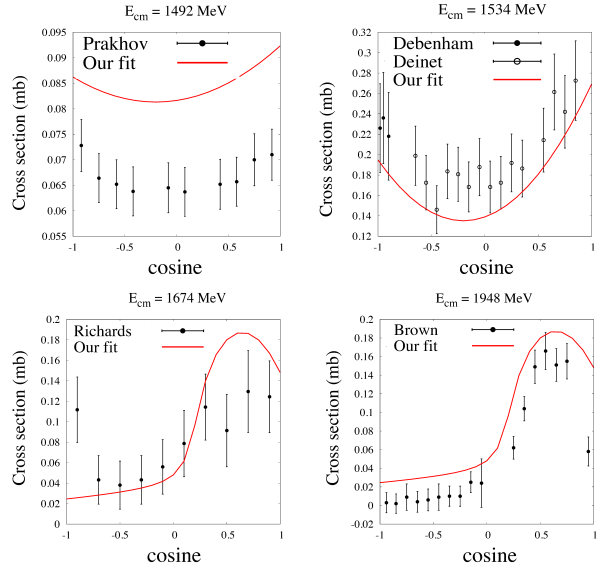}}
\caption{Cosine in the center of mass of the  $\eta$ produced in the $\pi^- p \rightarrow \eta n$ reaction. Our parametrization is given by the solid red line and the experimental data  are those of Richards {\it et al.} \cite{RIC70}, Prakhov {\it et al.} \cite{PRA05}, Brown {\it et al.} \cite{BRO79}, Debenham {\it et al.} \cite{DEB75} and Deinet {\it et al.} \cite{DEI69}.}
\label{Fig10}
\end{figure}

In this type of reaction only the direction of the emitted particles must be defined, energies being given by energy and momentum conservation. A parametrization of the cosine of the $\eta$ has been based on experimental data. It is given below and some examples are shown in fig.~\ref{Fig10}. The result obtained at low energy (left upper part of fig.~\ref{Fig10}) is assumed to be good enough, because at those energies emission is almost isotropic. It must be reminded that only the shape is relevant, since the cosine is drawn from the distribution.

Below $E_{cm}$ = 1650 MeV, the cosine distribution parametrization is $a_1\ cos^2\theta+b_1\ cos\theta+c_1$, with
\begin{eqnarray}\label{eq:ppiN2eta1}
a_1 =  &   2.5\ b_1 \nonumber\\
b_1 = & \frac{1}{2} \left( f_1 - \frac{f_1}{1.5 - 0.5\left(\frac{E_{cm}-1580}{95}\right)^2} \right) \nonumber\\
c_1 = & f_1 - 3.5\ b1 \nonumber\\
f_1 = & -2.88627\ 10^{-5}E_{cm}^2 + 9.155289\ 10^{-2}E_{cm} - 72.25436 \nonumber
\end{eqnarray}

Above $E_{cm}$ = 1650 MeV the parametrization is $(a_2\ cos^2\theta+b_2\ cos\theta+c_2) (0.5+\frac{arctan(10( cos\theta-0.2)))}{\pi} + 0.04$, with\\
$a_2 $= -0.29, $b_2 $= 0.348 and $c_2 $= 0.0546.

\medskip

\underline{NN $\rightarrow$ NN$\eta$+x$\pi$}
\medskip

Nucleon-Nucleon interactions can produce exclusive or inclusive (x =0 or $\ge$ 1) $\eta$ meson. The latter case has been explained in sect. \ref{XS-prod}. In all cases more than two particles exist in the final state and a phase-space generator is used to characterize each particle. This choice is exactly the same as the one used in the multiple pion case (NN $\rightarrow$ NN+y$\pi$, with y  $\ge$ 1). However, one already know that the use of a phase-space distribution is possibly not the best solution, as mentioned in the article of Vetter {\it et al.} \cite{VET91} where they compared the $\eta$ energy distributions in the reaction $pp\rightarrow pp\eta$ coming from a phase-space assumption and from an effective one-boson exchange model. Conclusions were that the phase-space distributions give more energy to the $\eta$ meson. This point must be kept in mind when analyzing calculation results.

Regarding the charge repartition, the procedures used for the multipion case can also be applied to the $\eta$ meson, because this latter has a spin equal to 0. Those procedures are based on isospin symmetries, G-parity and models, with assumptions when constraints were needed. This has been explained, in further detail, in \cite{PED11} and references therein.

Finally the case of NN $\rightarrow$ N$\Delta\eta$ is treated in the same way as NN $\rightarrow$ NN$\eta$, except that the mass of the $\Delta$ is chosen at random in a distribution as done for NN $\rightarrow$ N$\Delta$ \cite{CUG97}.

\subsubsection{Elastic scattering}
\label{Output-elas}
\medskip

\underline{$\eta$N $\rightarrow$ $\eta$N}
\medskip

\begin{figure}
\centering
\resizebox{0.58\textwidth}{!}{\includegraphics{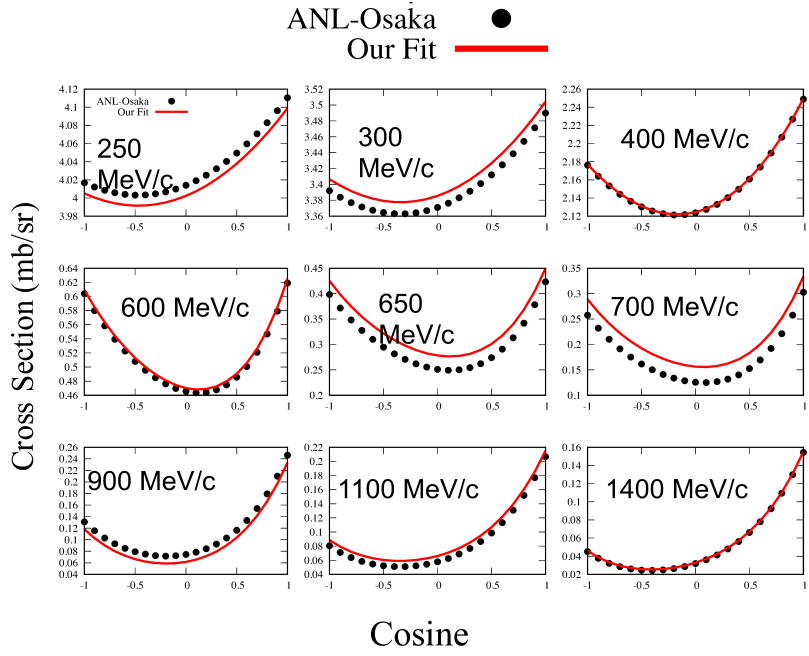}}
\caption{
Cosine in the center of mass of the outgoing $\eta$ in the elastic scattering $\eta$N $\rightarrow$ $\eta$N. Our parametrizations, given by the solid lines, are based on ANL-Osaka calculation results \cite{KAM13}, black marks.}
\label{Fig11}
\end{figure}

A parametrization of the cosine in the center of mass of the $\eta$ has been based on calculation results of the ANL-Osaka model \cite{KAM13}, already mentioned in previous sections. H. Kamano provided us with cosine distributions for several momenta of the $\eta$, up to 1400 MeV/c, from which a parametrization was obtained. Below an $\eta$ momentum of 250~MeV/c, emission is considered as isotropic, and above a polynomial form is used: \[\sum_{i=0}^6 a_i(P_{Lab})\ cos^i\theta\]
The $a_i(P_{Lab})$ are given in appendix \ref{Appelastic}, and some examples of cosine distribution shown in fig. \ref{Fig11} .

\subsubsection{Absorption}
\label{Output-abs}

\underline{$\eta$N $\rightarrow$ $\pi$N}
\medskip

\begin{figure}
\centering
\resizebox{0.58\textwidth}{!}{\includegraphics{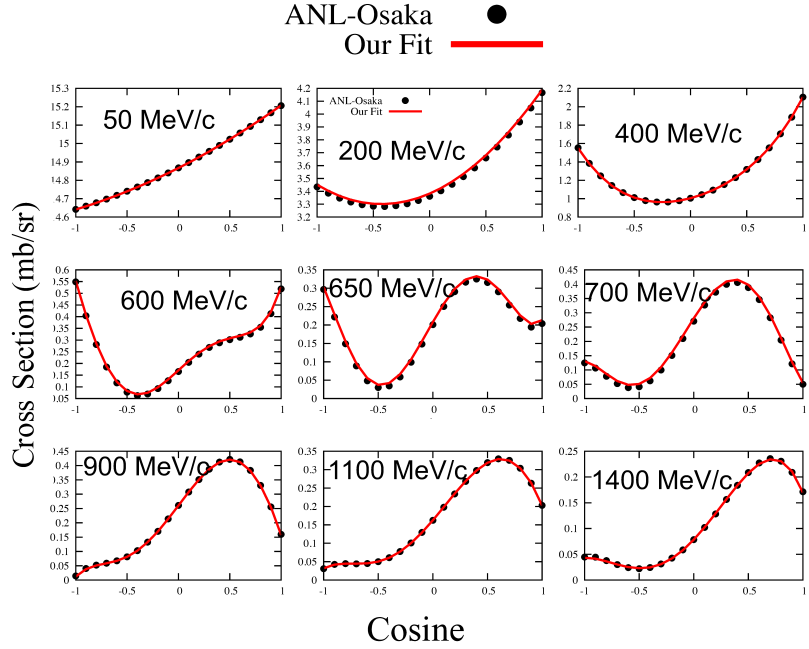}}
\caption{Cosine in the center of mass of the outgoing $\pi$ in the reaction $\eta$N $\rightarrow$ $\pi$N. Our parametrizations, given by the solid lines, are based on ANL-Osaka calculation results \cite{KAM13}, black marks.}
\label{Fig12}
\end{figure}

The $\pi$ cosine has been parametrized as the $\eta$ for the reaction $\eta$N $\rightarrow$ $\eta$N, here again thanks to calculations results from the ANL-Osaka model \cite{KAM13}. The only difference is that no isotropy was assumed below a given energy. Then a similar polynomial form was used and the parameters are given in appendix \ref{Appeta2pi} and some examples are shown in fig.~\ref{Fig12}.

\medskip

\underline{$\eta$N $\rightarrow$ $\pi$$\pi$N}
\medskip

Since in this case the final state is made of three particles, energies and directions are drawn from a phase-space generator.
The charges of pions and the nucleon are randomly chosen according to the probabilities coming from Clebsch-Gordan coefficients.

\subsection{Decays}
\label{Decays}

The decay channels taken into account for $\eta$ and $\omega$ are those with a branching ratio larger than 1\%. These branching ratios are taken from the Particle Data Group \cite{PDG14}, with obviously a renormalization to reach 100\%. Thus the channels implemented are $\gamma\gamma$~(39.72\%), $3\pi^0$~(32.93\%), $\pi^+\pi^-\pi^0$~(23.10\%) and $\pi^+\pi^-\gamma$~(4.25\%) for the $\eta$ meson, and  $\pi^+\pi^-\pi^0$~(90.09\%), $\pi^0\gamma$~(8.36\%) and $\pi^+\pi^-$~(1.55\%) for the $\omega$ meson. Isotropy is considered when two particles are emitted, otherwise a phase-space generator is used.

\subsection{Potentials}
\label{Potentials}

No consensus exists on the $\eta$-Nucleus potential. Numerous values are listed in a paper of Zhong {\it et al.} \cite{ZHO06} and they go from -26 MeV up to -88 MeV. Moreover a theoretical study \cite{NAG05} discusses the $\eta$ potential in the nucleus through chiral models, with values compatible with the ones previously mentioned (except in one case where this potential is repulsive in the core with an attractive part on the border), and with a dependence with the position in the nucleus. In the present version of INCL the dependence with the position is difficult to implement, therefore, as it will be shown in sect.~\ref{Results}, three values have been tested: 1.5 V$_{\pi^0}$, V$_{\pi^0}$ and 0 (where V$_{\pi^0}$~=~-30.6~MeV).
Finally a potential equal to zero has been chosen. This choice is explained in sect. \ref{Results}.

Concerning the $\omega$ meson, some values can be found in literature. However, unlike $\eta$, no experimental data helped us to decide which value seemed more suitable. Then the choice is up to now arbitrary and we assigned an attractive potential of -15 MeV (from \cite{FRI14}), even if other values are proposed, as -29 MeV from \cite{MET17}.

\section{Results and Discussion}
\label{Results}

\begin{figure}
\centering
\resizebox{0.52\textwidth}{!}{\includegraphics{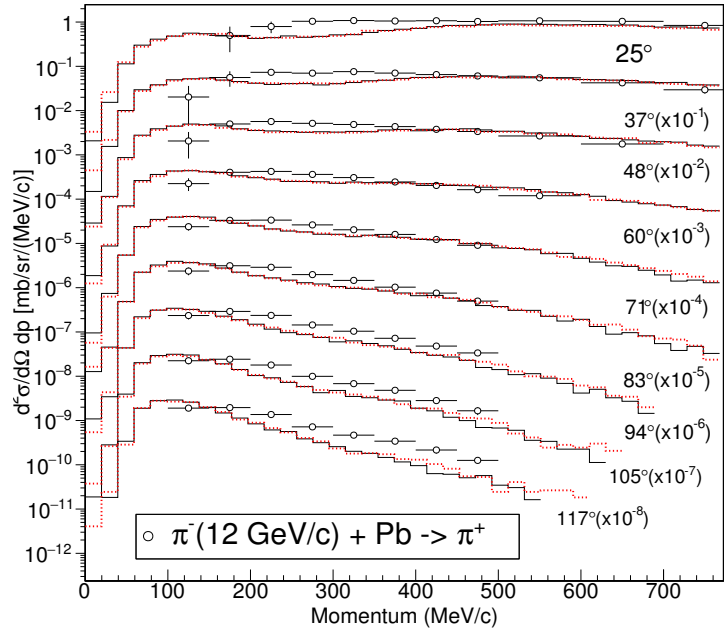}}
\caption{$\pi^+$ spectra with (dotted red lines) or without (solid black lines) $\eta$ and $\omega$ mesons in the intranuclear cascade code INCL. Experimental data are taken from \cite{CAT08} for the reaction $\pi^-$(12~GeV/c) + Pb $\rightarrow$ $\pi^+$ + X. A scaling factor (power of 10) is applied for each angle for sake of clarity.}
\label{Fig13}
\end{figure}

Available data concerning $\eta$ and $\omega$ mesons production from a nucleus hit by a light particle are scarce. While some measurements exist for the $\eta$, nothing has been published to our knowledge about the $\omega$. Before discussing $\eta$ production capabilities of our new version of INCL, fig.~\ref{Fig13} shows the impact of considering $\eta$ and $\omega$ on the $\pi$ production. Previous results of INCL \cite{MAN17} regarding the HARP measurements \cite{CAT08} were good, but with a deficiency in the 200 MeV/c - 400 MeV/c momentum region. 
With the decay of $\eta$ and $\omega$ in $\pi$ released latter, and so shielded from absorption in the nucleus, one aimed at testing if this could explain the deficiency in $\pi$ emission. The result is clear, $\eta$ and $\omega$ production has no impact on pion production.

\begin{figure}
\centering
\resizebox{0.45\textwidth}{!}{\subfloat{\label{Fig14a}\includegraphics{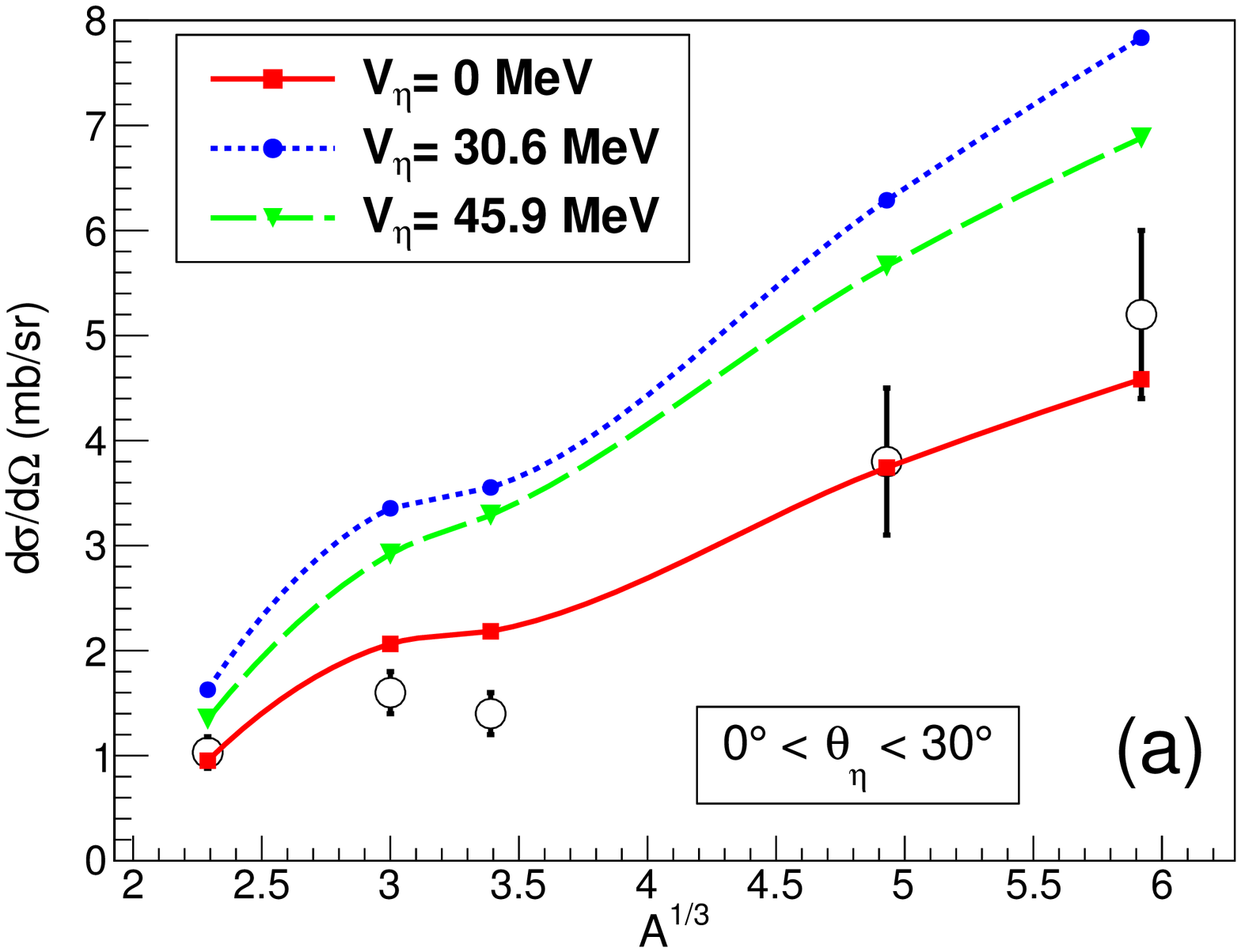}}}\\
\centering
\resizebox{0.45\textwidth}{!}{\subfloat{\label{Fig14b}\includegraphics{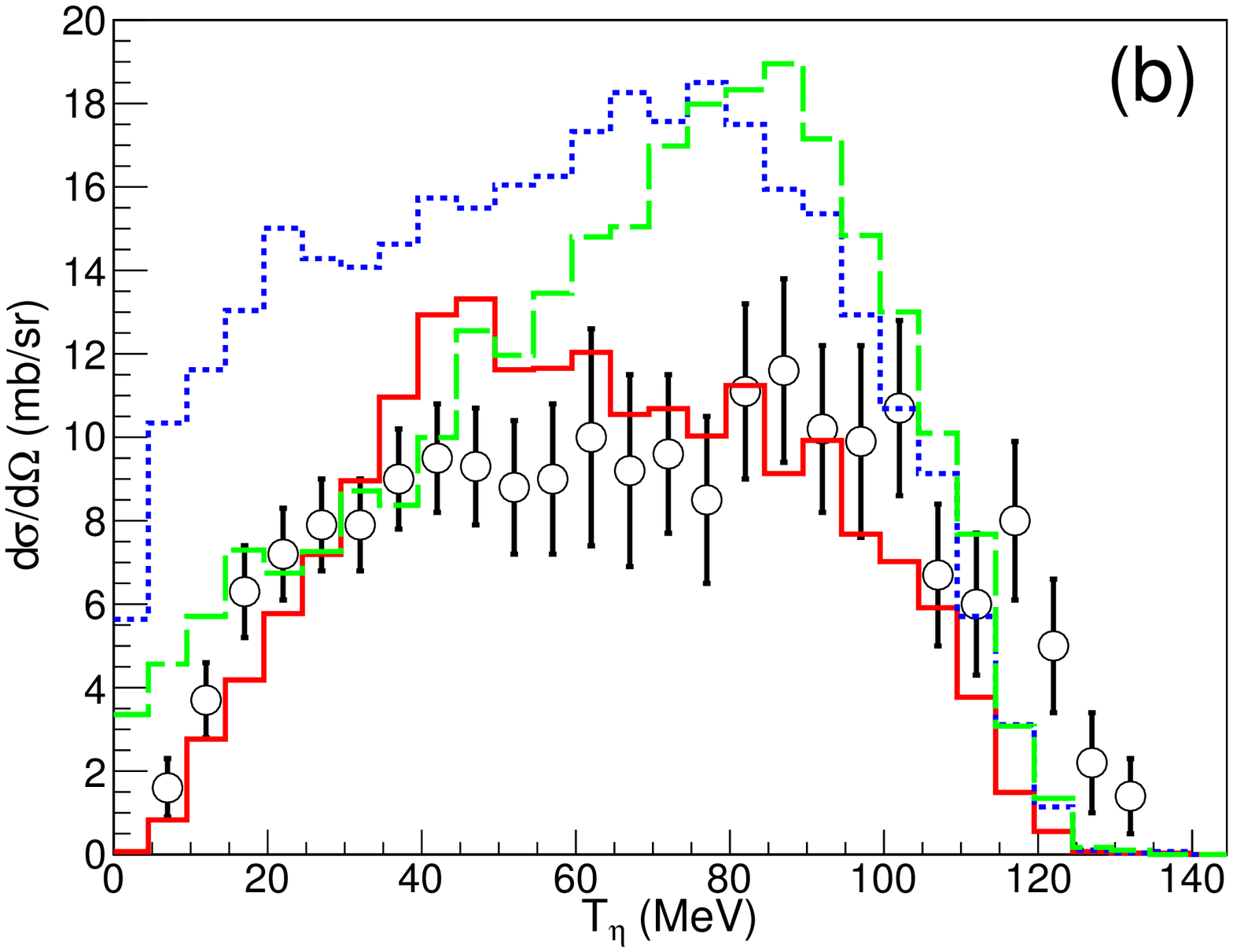}}}
\caption{ (a) A-dependence of the differential cross section of $\eta$-production in the reactions $\pi^+$(680 MeV/c)+nucleus (where nucleus stands from carbon to lead). (b) Energy spectrum of $\eta$ in the reaction $\pi^+$(680~MeV/c)+$^{12}C$. Experimental data (open black circles) come from \cite{GOL93}. Calculations are done by INCL with three different potentials for the $\eta$. More details in the text and on the figure.
\label{Fig14}
}
\end{figure}

Experimental data for only three cases of $\eta$ productions have been found and used to benchmark INCL. The first set deals with the production at threshold with the reaction $\pi^+$(680 MeV/c)+Nucleus \cite{GOL93}. Figure \ref{Fig14a} shows the production of $\eta$ for several targets from carbon to lead. In this low-energy case the potential felt in the nucleus by the $\eta$ plays a significant role in the production yield and the spectrum, thus three values has been tested: 0, -30.6 MeV (the $\pi^0$ potential used in INCL) and -45.9 (which is the $\pi^0$ potential multiplied by 1.5). For these data the best result is obtained with no $\eta$ potential. This result is possibly related to the place where the $\eta$'s are produced in the nucleus. Those measurements are performed in the forward direction ($\theta_\eta~<~$30\degree), then the detected $\eta$ comes probably from the border of the nucleus, undergoing few secondary scatterings. Therefore the value 0 for the potential is consistent with the predictions of \cite{NAG05}. Figure \ref{Fig14b} gives the $\eta$ spectrum in the forward direction ($\theta_\eta~<~$30\degree) for the reaction $\pi^+~$(680~MeV/c)~+~$^{12}C$. Here again the three values for the potential are shown and again the value 0 gives the best result.

\begin{figure}
\centering
\resizebox{0.45\textwidth}{!}{\subfloat{\label{Fig15a}\includegraphics{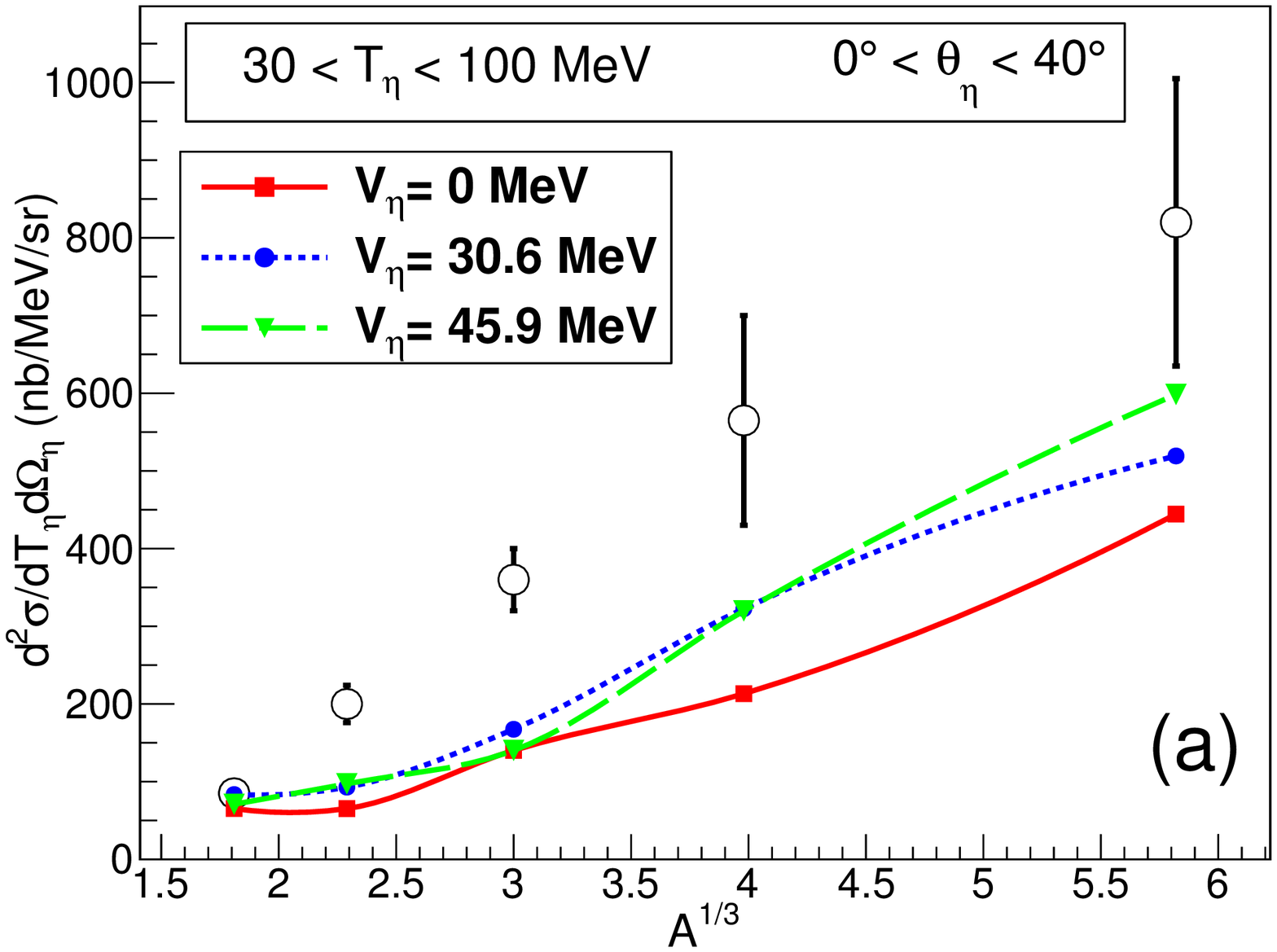}}}\\
\centering
\resizebox{0.45\textwidth}{!}{\subfloat{\label{Fig15b}\includegraphics{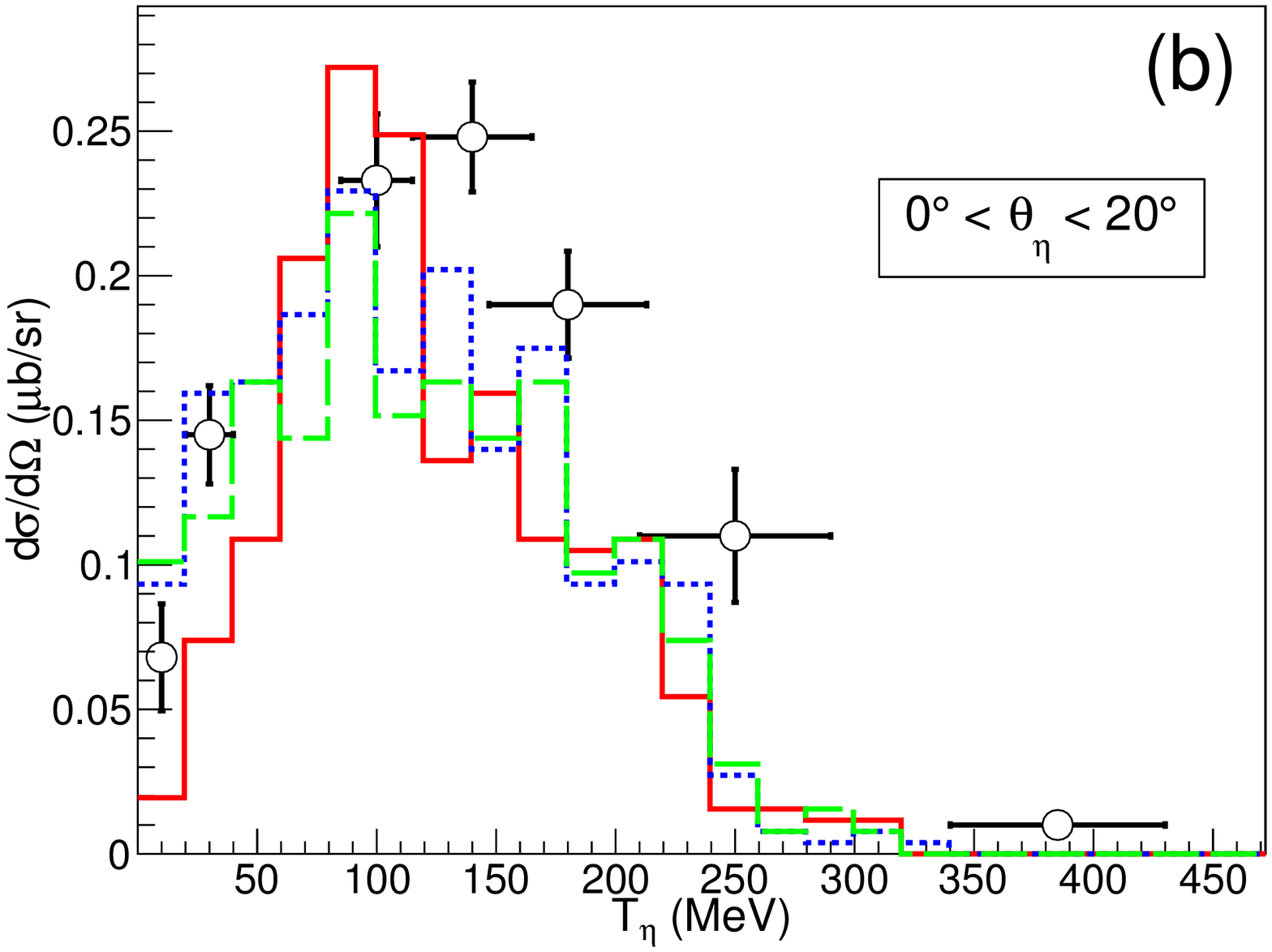}}}
\caption{ (a) A-dependence of the double-differential cross section of $\eta$-production in the reactions p(1 GeV)+nucleus (where nucleus stands from lithium to gold). (b) Energy spectrum of $\eta$ in the reaction p(1~GeV)+$^{11}B$. Experimental data (open black circles) come from \cite{GOL93}. Calculations are done by INCL with three different potentials for the $\eta$. More details in the text and on the figure.
\label{Fig15}
}
\end{figure}

The same kinds of experimental measurements with 1~GeV~proton-induced reactions have been used, but here at a subthreshold energy. Figure \ref{Fig15a} shows the production for different targets from lithium to gold, still in the forward direction ($\theta_\eta~<~$40\degree), but also with a cut-off in energy (30~MeV~$\le$~T$_\eta$~$\le$~100~MeV). The spectrum is shown in fig.~\ref{Fig15b}, with $\theta_\eta~<~$20\degree. In this case our calculation results are not too bad, but less good than in the previous case. In addition it is difficult to disentangle between the potentials. Actually, the projectile energy is below the $\eta$ production threshold, which makes the production mechanism strongly dependent on the Fermi momentum and then more tricky to analyse.

\begin{figure}
\centering
\resizebox{0.45\textwidth}{!}{\subfloat{\label{Fig16a}\includegraphics{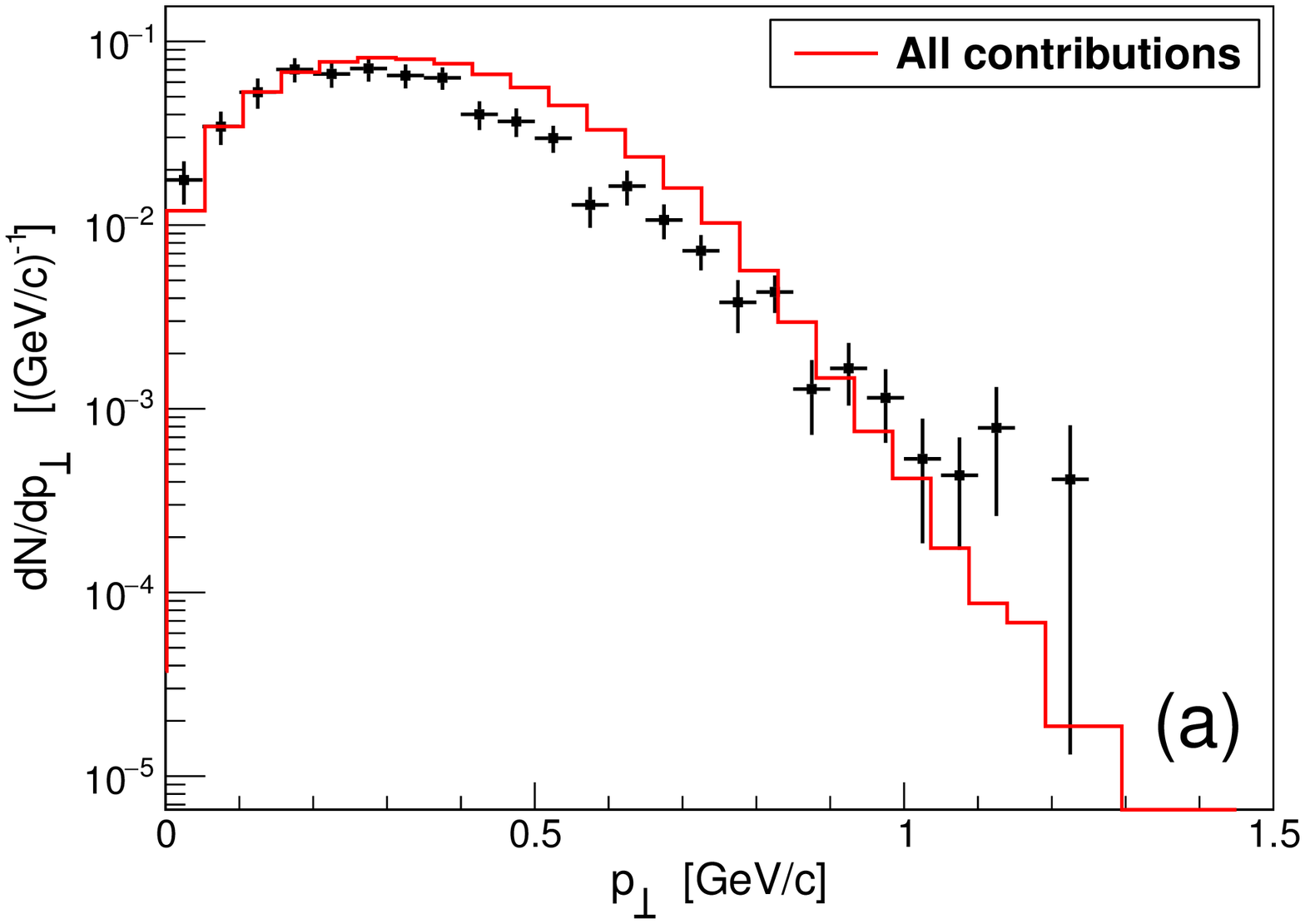}}}\\
\centering
\resizebox{0.45\textwidth}{!}{\subfloat{\label{Fig16b}\includegraphics{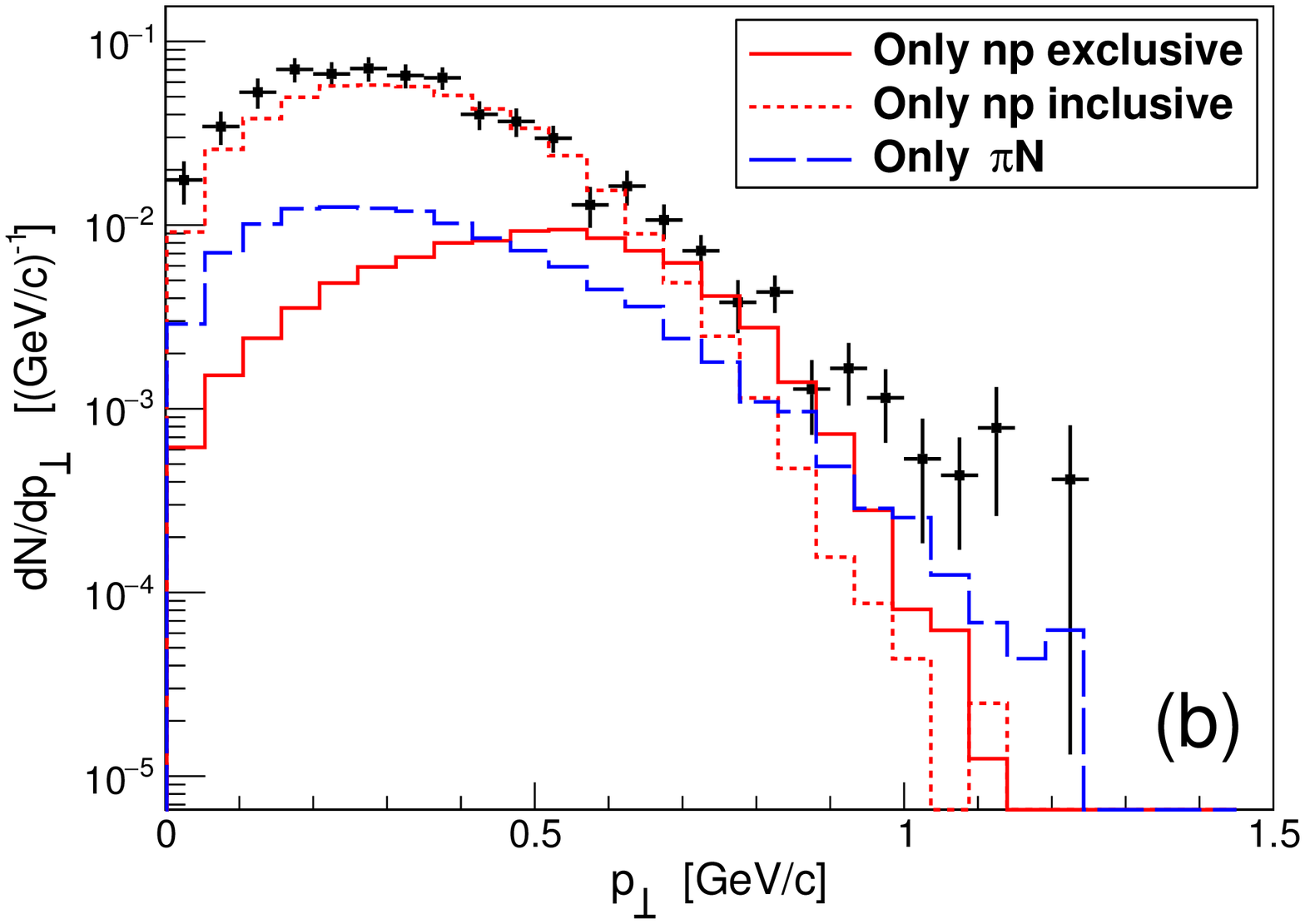}}}\\
\centering
\resizebox{0.45\textwidth}{!}{\subfloat{\label{Fig16c}\includegraphics{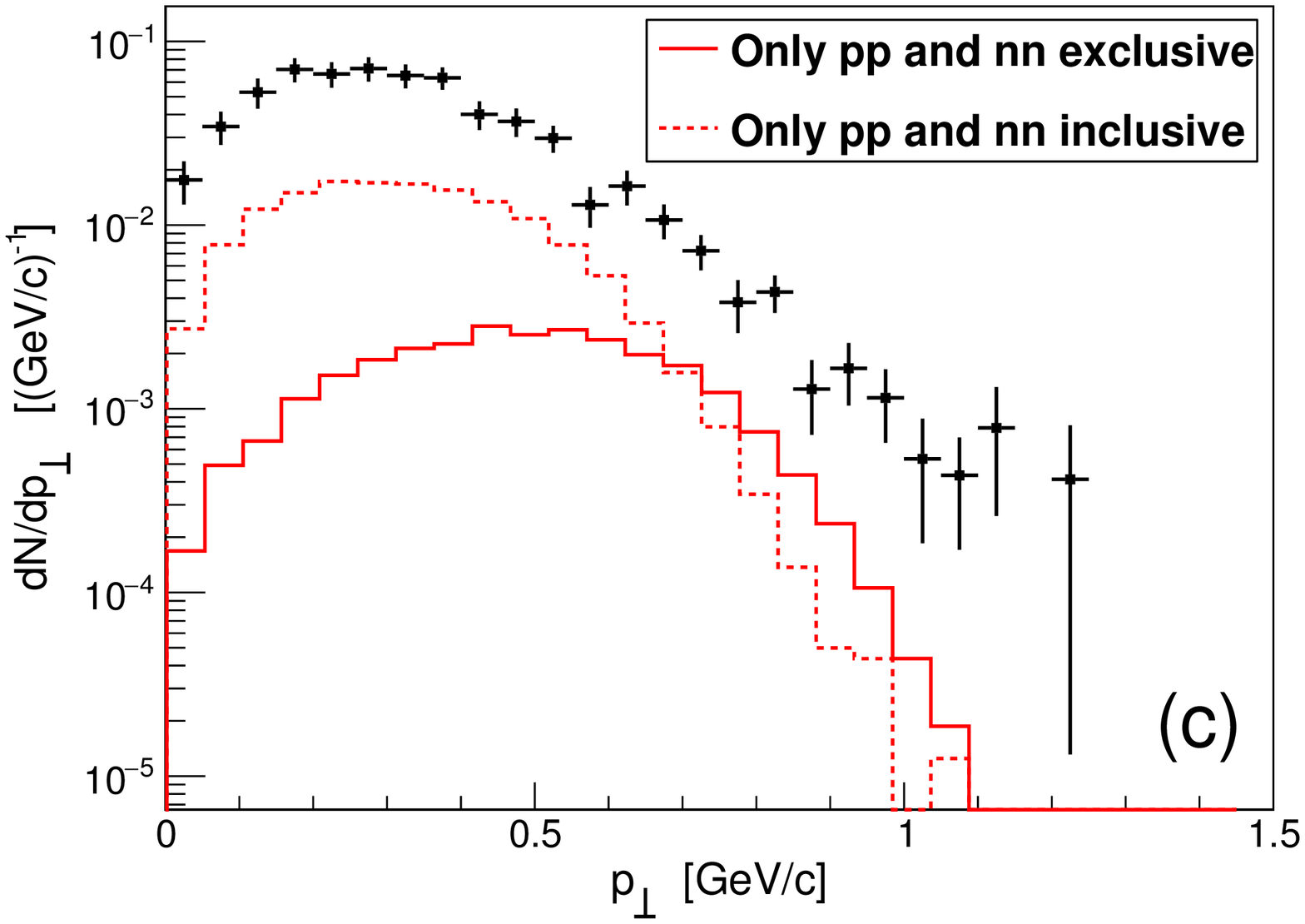}}}
\caption{ $\eta$ transverse momentum distributions dN/dp$_\perp$  in the p(3.5 GeV) + Nb reaction within the HADES rapidity range (0.2 $\le$ y $\le$ 1.8). INCL results are the lines and experimental data from \cite{AGA13}. (a) is the final result (all contributions), (b) contributions from elementary np and $\pi$N channels and (c) contributions from pp and nn channels. {\it Contribution} means a calculation result considering only a specific channel for $\eta$ production. This can explain, when the statistics is low (low yield), that a contribution can be higher than the sum of all contributions (a).
\label{Fig16}
}
\end{figure}
\begin{figure}
\centering
\resizebox{0.45\textwidth}{!}{\subfloat{\label{Fig17a}\includegraphics{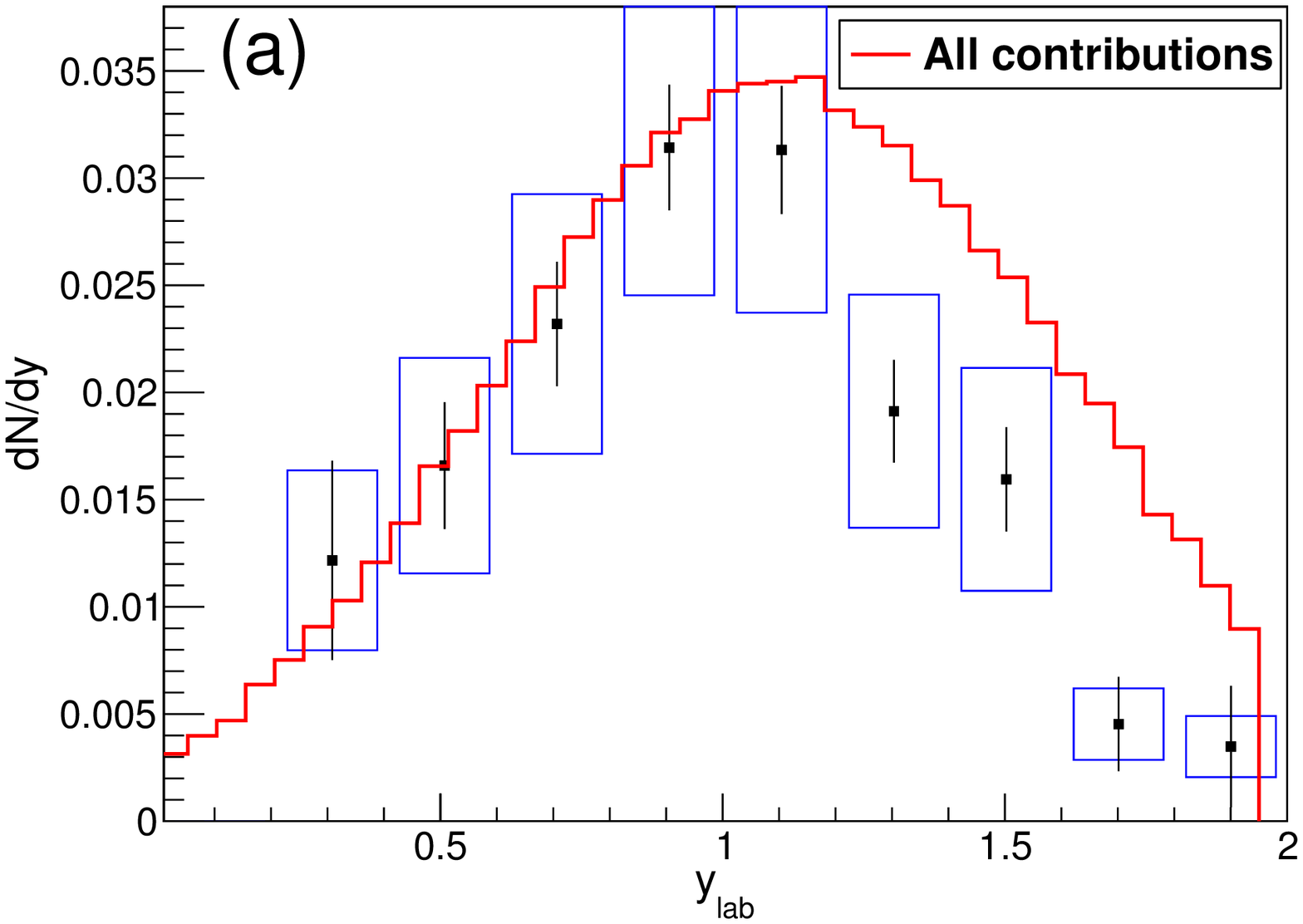}}}\\
\centering
\resizebox{0.45\textwidth}{!}{\subfloat{\label{Fig17b}\includegraphics{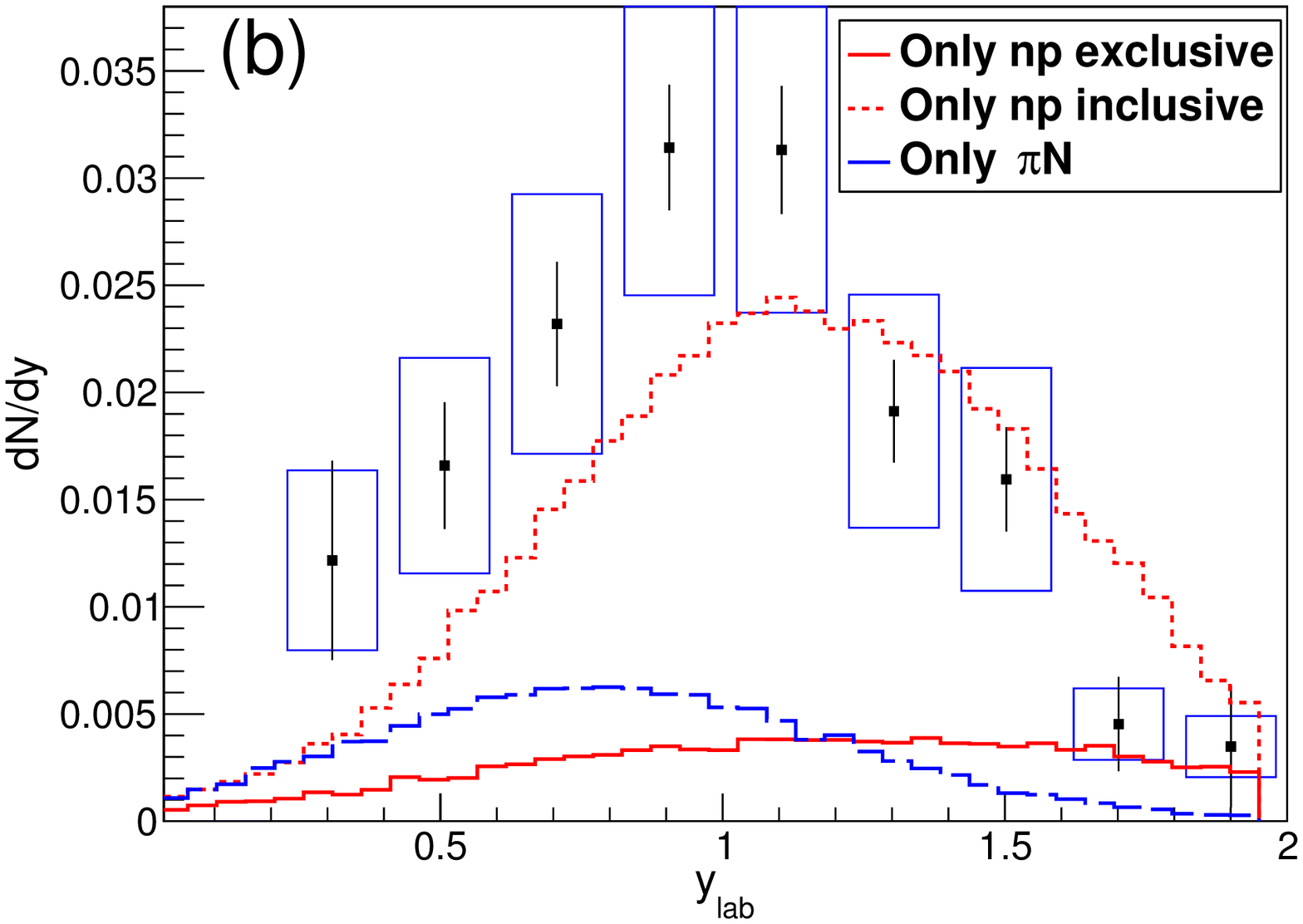}}}\\
\centering
\resizebox{0.45\textwidth}{!}{\subfloat{\label{Fig17c}\includegraphics{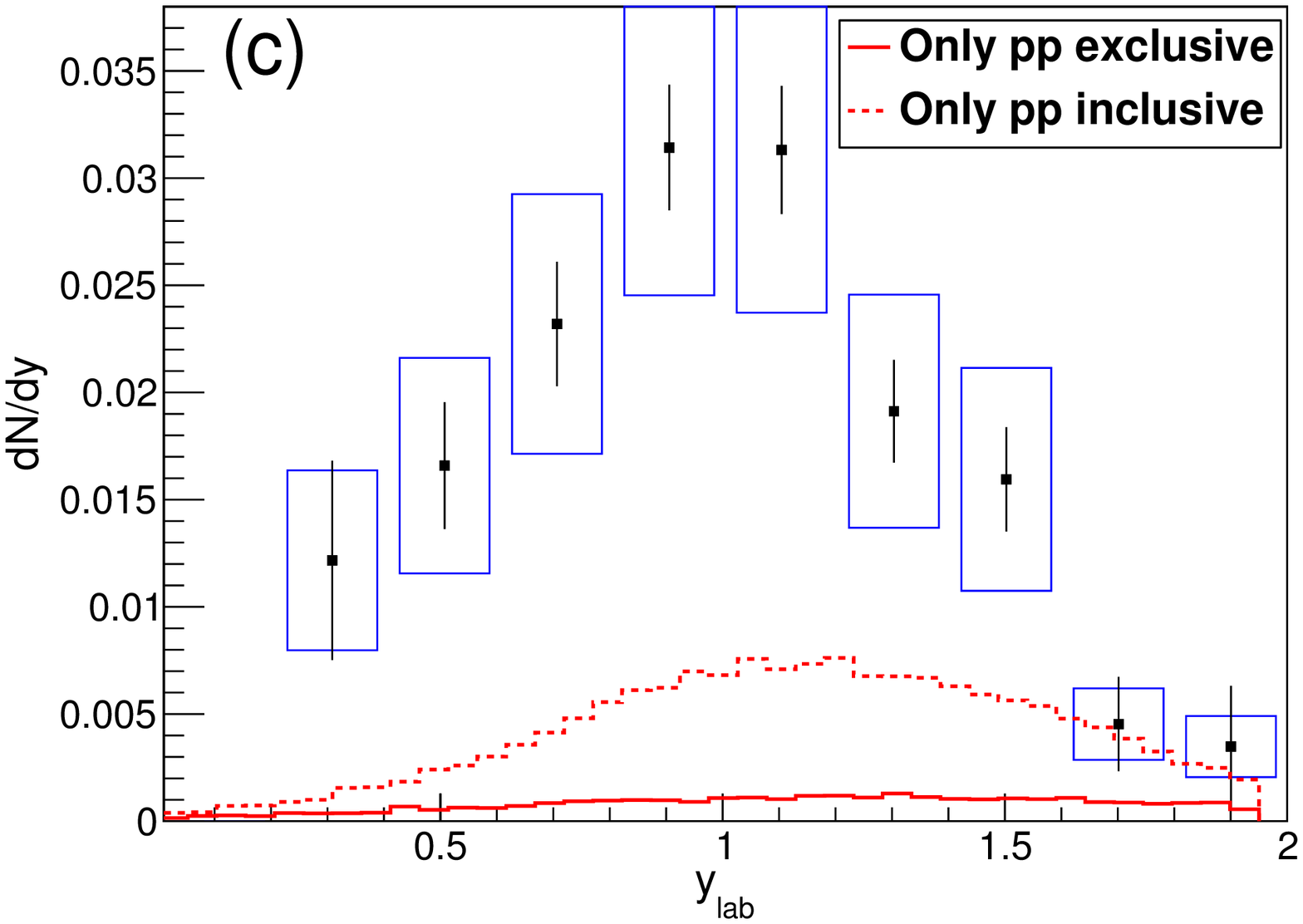}}}
\caption{ $\eta$ transverse momentum distributions dN/dp$_\perp$  in the p(3.5 GeV) + Nb reaction within the HADES rapidity range (0.2 $\le$ y $\le$ 1.8). INCL results are the lines and experimental data from \cite{AGA13} (blue boxes depict systematic errors). (a) is the final result (all contributions), (b) contributions from the elementary np and $\pi$N channels and (c) contributions from pp and nn channels. {\it Contribution} means a calculation result considering only a specific channel for $\eta$ production. This can explain, when the statistics is low (low yield), that a contribution can be higher than the sum of all contributions (a).
\label{Fig17}
}
\end{figure}

The last comparisons with experimental data deal with a HADES experiment where transverse momenta and rapidities of $\eta$'s were measured in the reaction p(3.5 GeV)~+~Nb \cite{AGA13}. Figure \ref{Fig16a} shows the transverse momenta obtained by our version of INCL and compared to the HADES measurements, as well as contributions of the elementary production processes. INCL fits well the experiment, with a slight overestimate of the momenta between 0.4 GeV and 0.8 GeV. For the highest momenta the experimental error bars are large and our statistics too low to say more than INCL give reasonable results even at those energies. It must be reminded here that the cross section of some elementary processes have been only roughly estimated, due to a lack of experimental data. This is especially the case for the production from np reactions, typically in the inclusive case, and this channel appears here as the main one, in particular at low momenta (fig. \ref{Fig16b}). The little overestimate, in intermediate energies, could come from the np reactions (exclusive and inclusive), which were estimated with difficulty in this energy range, as explained in sect. \ref{XS-prod} and as seen in fig. \ref{Fig3}. Moreover, as said in sect. \ref{Output-prod}, the energies of the $\eta$'s coming from nucleon-nucleon reactions are maybe too high, what could also explain the overestimate in the region around 0.7~GeV/c. On the contrary, the high energy part is due largely to the $\pi$N reactions (fig. \ref{Fig16b}) and in this case one can trust the cross sections used in INCL, mostly based on measurements. The pp and nn channels play a minor role (fig. \ref{Fig16c}), due to lower elementary cross sections. Those transverse momenta are in the end well simulated considering the uncertainties of the ingredients. This is true for the spectrum discussed just above, and also for the integrated inclusive $\eta$ multiplicity per p + Nb collision in the accepted rapidity range (0.2 $\le$ y $\le$ 1.8); The value calculated by INCL, 0.0384, is high but compatible with the one measured by HADES, 0.031 $\pm$ 0.002  $\pm$ 0.007.

Regarding the rapidity (y), the results shown in fig.~\ref{Fig17a} are once more correct. Here again contributions are displayed (fig.~\ref{Fig17b}-\ref{Fig17c}). While INCL matches the HADES measurements for the low rapidities (below y=1.2), it overestimates the highest ones (above y=1.2). This would indicate that the $\eta$'s are emitted too much in the forward direction. While at low energy there are two channels which contribute, {\it i.e.} $\pi$N and np inclusive, the high energy part is due to the np reactions (inclusive and exclusive) and inclusive pp. Here again it seems that the nucleon-nucleon reactions modeling can be improved and mainly for the energy given to the $\eta$, since this rapidity distribution corroborates, at least for the exclusive part, the results of Vetter \cite{VET91} showing that a pure phase-space distribution is not satisfactory to account for these data.

\section{Conclusion}
\label{Concl}

In this paper we discuss the new implementation of $\eta$ and $\omega$ mesons in the Li\`ege intranuclear cascade model (INCL). 
The first motivation was the study of the role of those particles with increasing energies . The second one was to enable INCL to simulate their production yields, especially for the $\eta$, whose rare decays are still studied.
Basic input ingredients are the elementary cross sections. As far as possible parametrizations are based on experimental data, but models are sometimes needed and in some places very large uncertainties remain. Those points are studied here to get a better analysis of the final calculation results and also to stress the need of new measurements. 

The production by $\pi$N interactions is based on experimental data and can be considered as reliable. Improvements could be done  beyond a pion energy of 2.7 GeV for the $\eta$ and of 12 GeV for the $\omega$ where we use extrapolations. The meson emission angle is based on experimental data for the $\eta$ up to a pion energy of $\sim$1.4~GeV, and our parametrization could lead to not enough forward peaked $\eta$'s for higher energies. An isotropic emission is assumed for the $\omega$. However, the main question about this production mode is the role and weight of the inclusive channel. It has not been taken into account due to a lack of information, but its role should increase with energy. The production by nucleon-nucleon interactions is less reliable than with $\pi$N interactions. 
Although in this case the inclusive channels are accounted for, we coped with several difficulties. First, those latter are only based on a model and only for the pp reaction. The np case is defined with hypotheses. Second, the exclusive channels are certainly based on experimental data, but, for the np case, only close to the threshold.
This is for the $\eta$, because for the $\omega$, if the same philosophy has been adopted, less refinements were considered, since unfortunately no experiment exists on the $\omega$ production from a nucleus. In addition, a phase-space generator is used to give the characteristics of the emitted mesons and probably, at least in the exclusive case, a dedicated model should give more realistic results. The elastic channel seems reasonnably parametrized, at least for the $\eta$, and should not be a problem. Absorption is considered only on the nucleon and one and two-pion emissions are the two channels accounted for here. Concerning the $\eta$, up to an incident energy of $\sim$900~MeV, cross sections are based on calculation results of a model, and beyond, the two-pion case is assumed to be equal to the one pion case, obtained thanks to the detailed balance. 
For the $\omega$ the detailed balance is used on the entire energy range for the one pion channel. The two-pion cross section is obtained by the difference between the total inelastic cross section (from a published parametrization) and the one pion channel. Here again there is room for improvements, especially when energy goes up.

While the input ingredients are not known as well as we could expect, INCL gives interesting calculation results on the $\eta$ production from a nucleus. Few experiments were performed on this type of observable and INCL was benchmarked on two experiments at low energy, close to the production threshold, and on one at higher energy. The low-energy experiments enable us to decide on the value of the in-medium potential felt by the $\eta$. The results were not bad at all with a potential equal to zero. This a priori curious value, could be due to the fact that, within the experiments considered here, the $\eta$ is produced at the surface of the nucleus. The HADES collaboration obtained data with a higher energy reaction, p(3.5~GeV)~+~Nb, and here again the results are rather good. One clear deficiency is that our $\eta$'s are too much emitted in the forward direction. This is possibly due to the phase-space generator used to define the characteristics of the final state when more than two particles are involved.

The implementation of the $\eta$ and $\omega$ mesons in an INC code suffers from lack of information regarding some elementary processes. Nevertheless, by using all available data drawn from experiments and models, the ingredients put in INCL allow to obtain good results compared to the scarce experimental data on $\eta$ production. Dedicated experiments both on elementary processes and on production from a nucleus are more than welcome. This INCL version with those two mesons is available in GEANT4.

\begin{acknowledgement}
The authors would like to thank  Bijan Saghai and  Hiroyuki Kamano for fruitfull discussions and calculation results regarding elementary processes ($\eta$), and the HADES collaboration for stimulating exchanges and especially B\' eatrice Ramstein, Malgorzata Gumberidze and Romain Holzmann for providing us their experimental data.
\end{acknowledgement}

\newpage

\appendix

\section{Polynomial parameters used in some cross sections parametrizations}
\label{xsparameters}
\subsection{$\pi^-$p $\rightarrow$ $\eta$n (eq. \ref{eq:piN2eta})\label{Apxspar1}}
\center
\begin{tabular}{p{1.2cm}p{1.2cm}p{1.2cm}p{1.2cm}p{1.2cm}p{1.2cm}p{1.2cm}}
& & & & & & \\
\multicolumn{7}{c}{}\\ \hline
\multicolumn{2}{p{2.4cm}}{$E_{cm}$\hspace{0.22 cm}$threshold$} & \multicolumn{2}{p{2.4cm}}{\hspace{1 cm}$1535.0$} & \multicolumn{2}{p{2.4cm}}{\hspace{1 cm}$1670.0$} & \multicolumn{1}{r@{\,}}{$1714.0$} \\ \hline\\\hline
\multicolumn{1}{|p{1.2cm}|}{$a_0$} & \multicolumn{2}{p{2.4cm}|}{$-2.03126590 \ 10^{6}$} & 
\multicolumn{2}{p{2.4cm}|}{$-2.28709420 \ 10^{5}$} & 
\multicolumn{2}{p{2.4cm}|}{$0$}  \\\hline 
\multicolumn{1}{|p{1.2cm}|}{$a_1$} & \multicolumn{2}{p{2.4cm}|}{$+5.30350527 \ 10^{3}$} & 
\multicolumn{2}{p{2.4cm}|}{$+5.67828367 \ 10^{2}$} & 
\multicolumn{2}{p{2.4cm}|}{$-5.66406200 \ 10^{-3}$}  \\\hline 
\multicolumn{1}{|p{1.2cm}|}{$a_2$} & \multicolumn{2}{p{2.4cm}|}{$-5.19310587$} & 
\multicolumn{2}{p{2.4cm}|}{$-5.28276144 \ 10^{-1}$} & 
\multicolumn{2}{p{2.4cm}|}{$+3.73776500 \ 10^{-6}$}  \\\hline 
\multicolumn{1}{|p{1.2cm}|}{$a_3$} & \multicolumn{2}{p{2.4cm}|}{$+2.26019390 \ 10^{-3}$} & 
\multicolumn{2}{p{2.4cm}|}{$+2.18279989 \ 10^{-4}$} & 
\multicolumn{2}{p{2.4cm}|}{$0$}  \\\hline 
\multicolumn{1}{|p{1.2cm}|}{$a_4$} & \multicolumn{2}{p{2.4cm}|}{$-3.68919797 \ 10^{-7}$} & 
\multicolumn{2}{p{2.4cm}|}{$-3.37986446 \ 10^{-8}$} & 
\multicolumn{2}{p{2.4cm}|}{$0$}  \\\hline 
& & & & & & \\
\end{tabular}

\fbox{$E_{cm}:$ MeV; $\sigma: mb$} 

\subsection{pp $\rightarrow$ pp$\eta$ (eq. \ref{eq:pp2ppeta})\label{Apxspar2}}
\center
\begin{tabular}{p{1.2cm}p{1.2cm}p{1.2cm}p{1.2cm}p{1.2cm}p{1.2cm}p{1.2cm}p{1.2cm}p{1.2cm}}
& & & & & & \\
\multicolumn{7}{c}{}\\ \hline
\multicolumn{2}{p{2.4cm}}{$E_{cm}$\hspace{0.2 cm}$threshold$} & \multicolumn{2}{p{2.4cm}}{\hspace{1 cm}$2.575$} & \multicolumn{2}{p{2.4cm}}{\hspace{1 cm}$2.725$} & \multicolumn{2}{p{2.4cm}}{\hspace{1 cm}$3.875$} & \multicolumn{1}{r@{\,}}{$$} \\ \hline\\\hline
\multicolumn{1}{|p{1.2cm}|}{$a_0$} & \multicolumn{2}{p{2.4cm}|}{$-5.97217435 \ 10^{6}$} & 
\multicolumn{2}{p{2.4cm}|}{$-2.0225 \ 10^{4}$} & 
\multicolumn{2}{p{2.4cm}|}{$+3.834747  \ 10^{5}$} & 
\multicolumn{2}{p{2.4cm}|}{$+3.6234   \ 10^{1}$}  \\\hline 
\multicolumn{1}{|p{1.2cm}|}{$a_1$} & \multicolumn{2}{p{2.4cm}|}{$+9.47729019 \ 10^{6}$} & 
\multicolumn{2}{p{2.4cm}|}{$+1.4692 \ 10^{4}$} & 
\multicolumn{2}{p{2.4cm}|}{$-5.824139 \ 10^{5}$} & 
\multicolumn{2}{p{2.4cm}|}{$+8.4531  \ 10^{1}$}  \\\hline 
\multicolumn{1}{|p{1.2cm}|}{$a_2$} & \multicolumn{2}{p{2.4cm}|}{$-5.63439990 \ 10^{6}$} & 
\multicolumn{2}{p{2.4cm}|}{$-2.6403 \ 10^{3}$} & 
\multicolumn{2}{p{2.4cm}|}{$+3.512942 \ 10^{5}$} & 
\multicolumn{2}{p{2.4cm}|}{$-1.3008  \ 10^{1}$}  \\\hline 
\multicolumn{1}{|p{1.2cm}|}{$a_3$} & \multicolumn{2}{p{2.4cm}|}{$+1.48722254 \ 10^{6}$} & 
\multicolumn{2}{p{2.4cm}|}{$0$} & 
\multicolumn{2}{p{2.4cm}|}{$-1.050549  \ 10^{5}$} & 
\multicolumn{2}{p{2.4cm}|}{$0$}  \\\hline 
\multicolumn{1}{|p{1.2cm}|}{$a_4$} & \multicolumn{2}{p{2.4cm}|}{$-1.47043497 \ 10^{5}$} & 
\multicolumn{2}{p{2.4cm}|}{$0$} & 
\multicolumn{2}{p{2.4cm}|}{$+1.556427   \ 10^{4}$} & 
\multicolumn{2}{p{2.4cm}|}{$0$}  \\\hline 
\multicolumn{1}{|p{1.2cm}|}{$a_5$} & \multicolumn{2}{p{2.4cm}|}{$0$} & 
\multicolumn{2}{p{2.4cm}|}{$0$} & 
\multicolumn{2}{p{2.4cm}|}{$-9.132809  \ 10^{2}$} & 
\multicolumn{2}{p{2.4cm}|}{$0$}  \\\hline 
& & & & & & \\
\end{tabular}

\fbox{$E_{cm}:$ GeV; $\sigma: \mu$b} 
\subsection{np $\rightarrow$ np$\eta$ (eq. \ref{eq:np2npeta})\label{Apxspar3}}
\center
\begin{tabular}{p{1.2cm}p{1.2cm}p{1.2cm}p{1.2cm}p{1.2cm}p{1.2cm}p{1.2cm}}
& & & & & & \\
\multicolumn{7}{c}{}\\ \hline
\multicolumn{2}{p{2.4cm}}{$E_{cm}$\hspace{0.2 cm}$threshold$} & \multicolumn{2}{p{2.4cm}}{\hspace{1 cm}$2.525$} & \multicolumn{2}{p{2.4cm}}{\hspace{1.2 cm}$3.5$} & \multicolumn{1}{r@{\,}}{$3.9$} \\ \hline\\\hline
\multicolumn{1}{|p{1.2cm}|}{$b_0$} & \multicolumn{2}{p{2.4cm}|}{$+1.02585558 \ 10^{5}$} & 
\multicolumn{2}{p{2.4cm}|}{$-4.510916 \ 10^{5}$} & 
\multicolumn{2}{p{2.4cm}|}{$+1.012000   \ 10^{5}$}  \\\hline 
\multicolumn{1}{|p{1.2cm}|}{$b_1$} & \multicolumn{2}{p{2.4cm}|}{$-8.49109854 \ 10^{4}$} & 
\multicolumn{2}{p{2.4cm}|}{$+5.716506 \ 10^{5}$} &
\multicolumn{2}{p{2.4cm}|}{$-8.082800  \ 10^{4}$}  \\\hline 
\multicolumn{1}{|p{1.2cm}|}{$b_2$} & \multicolumn{2}{p{2.4cm}|}{$+1.75702172 \ 10^{4}$} & 
\multicolumn{2}{p{2.4cm}|}{$-2.702126 \ 10^{5}$} &
\multicolumn{2}{p{2.4cm}|}{$+2.55600  \ 10^{4}$}  \\\hline 
\multicolumn{1}{|p{1.2cm}|}{$b_3$} & \multicolumn{2}{p{2.4cm}|}{$0$} & 
\multicolumn{2}{p{2.4cm}|}{$+5.658154  \ 10^{4}$} & 
\multicolumn{2}{p{2.4cm}|}{$-1.916200  \ 10^{3}$}  \\\hline 
\multicolumn{1}{|p{1.2cm}|}{$b_4$} & \multicolumn{2}{p{2.4cm}|}{$0$} & 
\multicolumn{2}{p{2.4cm}|}{$-4.433586   \ 10^{3}$} &
\multicolumn{2}{p{2.4cm}|}{$0$}  \\\hline 
& & & & & & \\
\end{tabular}

\fbox{$E_{cm}:$ GeV; $\sigma: \mu$b} 
\subsection{$\eta$N $\rightarrow$ $\eta$N (eq. \ref{eq:etaN2etaN})\label{Apxspar4}}
\center
\begin{tabular}{p{1.2cm}p{1.2cm}p{1.2cm}p{1.2cm}p{1.2cm}p{1.2cm}p{1.2cm}}
& & & & & & \\
\multicolumn{7}{c}{}\\ \hline
\multicolumn{2}{p{2.4cm}}{$P_{Lab}$\hspace{0.6 cm}$0$} & \multicolumn{2}{p{2.4cm}}{\hspace{1 cm}$700$} & \multicolumn{2}{p{2.4cm}}{\hspace{1 cm}$1400$} & \multicolumn{1}{r@{\,}}{$2025$} \\ \hline\\\hline
\multicolumn{1}{|p{1.2cm}|}{$a_0$} & \multicolumn{2}{p{2.4cm}|}{$+84.9650$} & 
\multicolumn{2}{p{2.4cm}|}{$+609.8501$} & 
\multicolumn{2}{p{2.4cm}|}{$+2.110529$}  \\\hline 
\multicolumn{1}{|p{1.2cm}|}{$a_1$} & \multicolumn{2}{p{2.4cm}|}{$-0.18379$} & 
\multicolumn{2}{p{2.4cm}|}{$-3.4092$} & 
\multicolumn{2}{p{2.4cm}|}{$-1.041950 \ 10^{-3}$}  \\\hline 
\multicolumn{1}{|p{1.2cm}|}{$a_2$} & \multicolumn{2}{p{2.4cm}|}{$+7.9188 \ 10^{-4}$} & 
\multicolumn{2}{p{2.4cm}|}{$+ 7.894845 \ 10^{-3}$} & 
\multicolumn{2}{p{2.4cm}|}{$0$}  \\\hline 
\multicolumn{1}{|p{1.2cm}|}{$a_3$} & \multicolumn{2}{p{2.4cm}|}{$- 4.3222 \ 10^{-6}$} & 
\multicolumn{2}{p{2.4cm}|}{$- 9.667078 \ 10^{-6}$} & 
\multicolumn{2}{p{2.4cm}|}{$0$}  \\\hline 
\multicolumn{1}{|p{1.2cm}|}{$a_4$} & \multicolumn{2}{p{2.4cm}|}{$+9.7914 \ 10^{-9}$} & 
\multicolumn{2}{p{2.4cm}|}{$+6.601312 \ 10^{-9}$} & 
\multicolumn{2}{p{2.4cm}|}{$0$}  \\\hline 
\multicolumn{1}{|p{1.2cm}|}{$a_5$} & \multicolumn{2}{p{2.4cm}|}{$-9.7815 \ 10^{-12}$} & 
\multicolumn{2}{p{2.4cm}|}{$-2.384766 \ 10^{-12}$} & 
\multicolumn{2}{p{2.4cm}|}{$0$}  \\\hline 
\multicolumn{1}{|p{1.2cm}|}{$a_6$} & \multicolumn{2}{p{2.4cm}|}{$+3.6838 \ 10^{-15}$} & 
\multicolumn{2}{p{2.4cm}|}{$+3.56263 \ 10^{-16}$} & 
\multicolumn{2}{p{2.4cm}|}{$0$}  \\\hline 
& & & & & & \\
\end{tabular}

\fbox{$P_{Lab}:$ MeV/c; $\sigma: mb$} 

\subsection{$\eta$N $\rightarrow$ $\pi$N (eq. \ref{eq:etaN2piN})\label{Apxspar5}}
\center
\begin{tabular}{p{1.2cm}p{1.2cm}p{1.2cm}p{1.2cm}p{1.2cm}p{1.2cm}p{1.2cm}}
& & & & & & \\
\multicolumn{7}{c}{}\\ \hline
\multicolumn{2}{p{2.4cm}}{$P_{Lab}$\hspace{0.6 cm}$0$} & \multicolumn{2}{p{2.4cm}}{\hspace{1 cm}$574$} & \multicolumn{2}{p{2.4cm}}{\hspace{1 cm}$850$} & \multicolumn{1}{r@{\,}}{$1300$} \\ \hline\\ \hline
\multicolumn{1}{|p{1.2cm}|}{$a_0$} & \multicolumn{2}{p{2.4cm}|}{$+403.1449$} & 
\multicolumn{2}{p{2.4cm}|}{$-10694.3$} & 
\multicolumn{2}{p{2.4cm}|}{$-1.70427$}  \\\hline 
\multicolumn{1}{|p{1.2cm}|}{$a_1$} & \multicolumn{2}{p{2.4cm}|}{$-6.172108$} & 
\multicolumn{2}{p{2.4cm}|}{$+91.8098$} &
\multicolumn{2}{p{2.4cm}|}{$+1.84148 \ 10^{-2}$}  \\\hline 
\multicolumn{1}{|p{1.2cm}|}{$a_2$} & \multicolumn{2}{p{2.4cm}|}{$+4.437913 \ 10^{-2}$} & 
\multicolumn{2}{p{2.4cm}|}{$-3.24936 \ 10^{-1}$} & 
\multicolumn{2}{p{2.4cm}|}{$-2.07653 \ 10^{-5}$}  \\\hline 
\multicolumn{1}{|p{1.2cm}|}{$a_3$} & \multicolumn{2}{p{2.4cm}|}{$-1.681980 \ 10^{-4}$} & 
\multicolumn{2}{p{2.4cm}|}{$+6.07658 \ 10^{-4}$} & 
\multicolumn{2}{p{2.4cm}|}{$+6.56364 \ 10^{-9}$}  \\\hline 
\multicolumn{1}{|p{1.2cm}|}{$a_4$} & \multicolumn{2}{p{2.4cm}|}{$+3.443487 \ 10^{-7}$} & 
\multicolumn{2}{p{2.4cm}|}{$-6.33891 \ 10^{-7}$} & 
\multicolumn{2}{p{2.4cm}|}{$0$}  \\\hline 
\multicolumn{1}{|p{1.2cm}|}{$a_5$} & \multicolumn{2}{p{2.4cm}|}{$-3.603636 \ 10^{-10}$} & 
\multicolumn{2}{p{2.4cm}|}{$+3.50041 \ 10^{-10}$} &
\multicolumn{2}{p{2.4cm}|}{$0$}  \\\hline 
\multicolumn{1}{|p{1.2cm}|}{$a_6$} & \multicolumn{2}{p{2.4cm}|}{$+1.511147 \ 10^{-13}$} & 
\multicolumn{2}{p{2.4cm}|}{$-8.00018 \ 10^{-14}$} &
\multicolumn{2}{p{2.4cm}|}{$0$}  \\\hline 
& & & & & & \\
\end{tabular}

\fbox{$P_{Lab}:$ MeV/c; $\sigma: mb$} 

\subsection{$\eta$N $\rightarrow$ $\pi\pi$N (eq. \ref{eq:etaN2pipiN})\label{Apxspar6}}
\center
\begin{tabular}{p{1.2cm}p{1.2cm}p{1.2cm}p{1.2cm}p{1.2cm}p{1.2cm}p{1.2cm}}
& & & & & & \\
\multicolumn{7}{c}{}\\ \hline
\multicolumn{2}{p{2.4cm}}{$P_{Lab}$\hspace{0.6 cm}$0$} & \multicolumn{2}{p{2.4cm}}{\hspace{1 cm}$450$} & \multicolumn{2}{p{2.4cm}}{\hspace{1.8 cm}$600$} & \multicolumn{1}{r@{\,}}{$1300$} \\ \hline\\ \hline
\multicolumn{1}{|p{1.2cm}|}{$b_0$} & \multicolumn{2}{p{2.4cm}|}{$+110.358550$} & 
\multicolumn{2}{p{2.4cm}|}{} & 
\multicolumn{2}{p{2.4cm}|}{$-609.447145$}  \\\cline{1-3}\cline{6-7} 
\multicolumn{1}{|p{1.2cm}|}{$b_1$} & \multicolumn{2}{p{2.4cm}|}{$-2.03113098$} & 
\multicolumn{2}{p{2.4cm}|}{} &
\multicolumn{2}{p{2.4cm}|}{$+3.93492126$}  \\\cline{1-3}\cline{6-7} 
\multicolumn{1}{|p{1.2cm}|}{$b_2$} & \multicolumn{2}{p{2.4cm}|}{$+1.83382964 \ 10^{-2}$} & 
\multicolumn{2}{p{2.4cm}|}{} & 
\multicolumn{2}{p{2.4cm}|}{$-1.01830486 \ 10^{-2}$}  \\\cline{1-3}\cline{6-7} 
\multicolumn{1}{|p{1.2cm}|}{$b_3$} & \multicolumn{2}{p{2.4cm}|}{$-9.01422901 \ 10^{-5}$} & 
\multicolumn{2}{c|}{$\sigma$($P_{Lab}$  =  450  MeV/c)} & 
\multicolumn{2}{p{2.4cm}|}{$+1.37055547 \ 10^{-5}$}  \\\cline{1-3}\cline{6-7} 
\multicolumn{1}{|p{1.2cm}|}{$b_4$} & \multicolumn{2}{p{2.4cm}|}{$+2.46011585 \ 10^{-7}$} & 
\multicolumn{2}{p{2.4cm}|}{} & 
\multicolumn{2}{p{2.4cm}|}{$- 1.01727714 \ 10^{-7}$}  \\\cline{1-3}\cline{6-7}  
\multicolumn{1}{|p{1.2cm}|}{$b_5$} & \multicolumn{2}{p{2.4cm}|}{$-3.49750459 \ 10^{-10}$} & 
\multicolumn{2}{p{2.4cm}|}{} &
\multicolumn{2}{p{2.4cm}|}{$+3.95985900 \ 10^{-12}$}  \\\cline{1-3}\cline{6-7}  
\multicolumn{1}{|p{1.2cm}|}{$b_6$} & \multicolumn{2}{p{2.4cm}|}{$+2.01854221 \ 10^{-13}$} & 
\multicolumn{2}{p{2.4cm}|}{} &
\multicolumn{2}{p{2.4cm}|}{$-6.32793049 \ 10^{-16}$}  \\\hline 
& & & & & & \\
\end{tabular}

\fbox{$P_{Lab}:$ MeV/c; $\sigma: mb$}

\vspace{-0.3 cm}
\section{Parameters $a_i$ of the $\eta$ cosine distribution in the elastic channel ($\eta$N $\rightarrow$ $\eta$N)}
\label{Appelastic}
\center
$f(cos\theta) = \sum_{i=0}^6 \mathbf{a_i}(P_{Lab})\ cos^i\theta,$\hspace{1cm} with\hspace{1cm}  $\mathbf{a_i}(P_{Lab})=\sum_{i=0}^6 b_i\ P_{Lab}^i$
\medskip

\fbox{$P_{Lab}:$ MeV/c; $\frac{d\sigma}{d\Omega}: mb/sr$} 
\center
\begin{tabular}{|c|c||c|c|c|c|}\hline
$\mathbf{a_0}$  & $\forall P_{Lab}$                                   & $\mathbf{a_1}$ & $P_{Lab} < 500.$ & $500 \le P_{Lab} < 750.$ & $ 750. \le P_{Lab}$   \\\hline\hline
$b_0$                &$+2.609971\hspace{0.2cm} $               & $b_0$                &$+8.701280\hspace{0.2cm} 10^{-4}$   & $-4.152037\hspace{0.2cm} $                & $-5.030932\hspace{0.2cm} 10^{-2} $  \\\hline
$b_1$                &$+3.856266\hspace{0.2cm} 10^{-2}$   & $b_1$                &$-3.001598\hspace{0.2cm} 10^{-4}$    & $+2.917630\hspace{0.2cm} 10^{-2}$   & $-3.318304\hspace{0.2cm} 10^{-3}$   \\\hline
$b_2$                &$-2.147259\hspace{0.2cm} 10^{-4}$    & $b_2$               &$+5.607250\hspace{0.2cm} 10^{-6}$   & $-7.447474\hspace{0.2cm} 10^{-5}$    & $+1.347462\hspace{0.2cm} 10^{-5}$   \\\hline
$b_3$                &$+4.181510\hspace{0.2cm} 10^{-7}$   & $b_3$                &$-2.129570\hspace{0.2cm} 10^{-8}$    & $+8.168681\hspace{0.2cm} 10^{-8}$   & $-2.107063\hspace{0.2cm} 10^{-8}$   \\\hline
$b_4$                &$-3.912863\hspace{0.2cm} 10^{-10}$  & $b_4$                &$+3.007021\hspace{0.2cm} 10^{-11}$ & $-3.255396\hspace{0.2cm} 10^{-11}$  & $+1.638691\hspace{0.2cm} 10^{-11}$ \\\hline
$b_5$                &$+1.789654\hspace{0.2cm} 10^{-13}$ & $b_5$                &$-1.524408\hspace{0.2cm} 10^{-14}$  & $0$                                                        & $-6.380168\hspace{0.2cm} 10^{-15}$ \\\hline
$b_6$                &$-3.220143\hspace{0.2cm} 10^{-17}$  & $b_6$                &$0$                                                       & $0$                                                         & $+9.964504\hspace{0.2cm} 10^{-19}$\\\hline
\multicolumn{3}{c}{}\\ 
\end{tabular}
\begin{tabular}{|c|c|c|c||c|c|c|}\hline
$\mathbf{a_2}$    & $P_{Lab} < 500.$                                  & $500 \le P_{Lab} < 750.$                       & $ 750. \le P_{Lab}$                                & $\mathbf{a_3}$    & $P_{Lab} < 650.$ & $650 \le P_{Lab} $   \\\hline\hline
$b_0$                  &$+4.58496\hspace{0.2cm} 10^{-1}$     & $+1.309433\hspace{0.2cm} 10^{+1}$   & $-6.025497\hspace{0.2cm} $               &$b_0$                    & $+2.832772\hspace{0.2cm} 10^{-1}$   & $-5.624556\hspace{0.2cm} $               \\\hline
$b_1$                  &$-7.218145\hspace{0.2cm} 10^{-3}$    & $-8.742722\hspace{0.2cm} 10^{-2}$    & $+3.652772\hspace{0.2cm} 10^{-2}$   &$b_1$                   &  $-4.245566\hspace{0.2cm} 10^{-3}$    & $+3.500692\hspace{0.2cm} 10^{-2}$\\\hline
$b_2$                  &$+4.292106\hspace{0.2cm} 10^{-5}$   & $+2.171883\hspace{0.2cm} 10^{-4}$    & $-8.778573\hspace{0.2cm} 10^{-5}$   &$b_2$                    & $+2.572396\hspace{0.2cm} 10^{-5}$   & $-8.812510\hspace{0.2cm} 10^{-5}$ \\\hline
$b_3$                  &$-1.124158\hspace{0.2cm} 10^{-7}$    & $-2.362724\hspace{0.2cm} 10^{-7}$    & $+1.100124\hspace{0.2cm} 10^{-7}$   &$b_3$                   &  $-8.036891\hspace{0.2cm} 10^{-8}$    & $+1.159487\hspace{0.2cm} 10^{-8}$ \\\hline
$b_4$                  &$+1.354078\hspace{0.2cm} 10^{-10}$ & $+9.512730\hspace{0.2cm} 10^{-11}$  & $-7.640831\hspace{0.2cm} 10^{-11}$  &$b_4$                   &  $+1.357165\hspace{0.2cm} 10^{-10}$ & $-8.435635\hspace{0.2cm} 10^{-11}$ \\\hline
$b_5$                  &$-6.085067\hspace{0.2cm} 10^{-14}$  & $0$                                                        & $+2.798222\hspace{0.2cm} 10^{-14}$ &$b_5$                   &  $-1.151454\hspace{0.2cm} 10^{-13}$  & $+3.223757\hspace{0.2cm} 10^{-14}$\\\hline
$b_6$                  &$0$                                                        & $0$                                                        & $-4.228889\hspace{0.2cm} 10^{-18}$  &$b_6$                   &  $+3.783071\hspace{0.2cm} 10^{-17}$ & $-5.063316\hspace{0.2cm} 10^{-18 }$\\\hline
\multicolumn{3}{c}{}\\
\end{tabular}
\begin{tabular}{|c|c|c||c|c|c|c|}\hline
$\mathbf{a_4}$   & $P_{Lab} < 700.$                                  & $700 \le P_{Lab} $                                & $\mathbf{a_5}$    & $P_{Lab} < 650.$                               & $650 \le P_{Lab} < 950.$                         & $ 950. \le P_{Lab}$  \\\hline\hline
$b_0$                 &$+4.684685\hspace{0.2cm} 10^{-1}$   & $-5.237677\hspace{0.2cm} $               & $b_0$                  &$-2.969608\hspace{0.2cm} 10^{-2}$ & $-7.065712\hspace{0.2cm} 10^{-1}$      & $+6.534893\hspace{0.2cm} 10^{-1}$ \\\hline
$b_1$                 &$-6.822100\hspace{0.2cm} 10^{-3}$    & $+3.029285\hspace{0.2cm} 10^{-2}$  & $b_1$                  &$+4.761016\hspace{0.2cm} 10^{-4}$ & $+2.146666\hspace{0.2cm} 10^{-3}$     & $-3.205628\hspace{0.2cm} 10^{-3}$\\\hline
$b_2$                 &$+3.988902\hspace{0.2cm} 10^{-5}$   & $-7.113554\hspace{0.2cm} 10^{-5}$   & $b_2$                  &$-3.150857\hspace{0.2cm} 10^{-6}$ & $+2.227237\hspace{0.2cm} 10^{-6}$     & $+6.604074\hspace{0.2cm} 10^{-6}$ \\\hline
$b_3$                 &$-1.192317\hspace{0.2cm} 10^{-7}$    & $+8.772790\hspace{0.2cm} 10^{-8}$  & $b_3$                  &$+1.100580\hspace{0.2cm} 10^{-8}$ & $-1.678272\hspace{0.2cm} 10^{-8}$     & $-7.303856\hspace{0.2cm} 10^{-9}$ \\\hline
$b_4$                 &$+1.907868\hspace{0.2cm} 10^{-10}$ & $-6.012288\hspace{0.2cm} 10^{-11}$ & $b_4$                  &$-2.136095\hspace{0.2cm} 10^{-11}$ & $+2.625428\hspace{0.2cm} 10^{-11}$ & $+4.578142\hspace{0.2cm} 10^{-12}$ \\\hline
$b_5$                 &$-1.534471\hspace{0.2cm} 10^{-13}$  & $+2.174395\hspace{0.2cm} 10^{-14}$& $b_5$                  &$+2.176771\hspace{0.2cm} 10^{-14}$ & $-1.756295\hspace{0.2cm} 10^{-14}$ & $-1.546647\hspace{0.2cm} 10^{-15}$ \\\hline
$b_6$                 &$+4.826684\hspace{0.2cm} 10^{-17}$ & $-3.245143\hspace{0.2cm} 10^{-18 }$& $b_6$                  &$-9.021076\hspace{0.2cm} 10^{-18}$ & $+4.424756\hspace{0.2cm} 10^{-18}$ & $+2.209585\hspace{0.2cm} 10^{-19}$  \\\hline
\multicolumn{3}{c}{}\\ 
\end{tabular}
\begin{tabular}{|c|c|c|c|}\hline
$\mathbf{a_6}$    & $P_{Lab} < 300.$ & $300 \le P_{Lab} < 500.$ & $ 500. \le P_{Lab}$  \\\hline\hline
$b_0$ &$-1.1545210\hspace{0.2cm} 10^{-4}$ & $+3.622575\hspace{0.2cm} 10^{-3}$   & $-1.443857\hspace{0.2cm} 10^{-3}$  \\\hline
$b_1$ &$-8.3834000\hspace{0.2cm} 10^{-8}$ & $-3.986627\hspace{0.2cm} 10^{-5}$   & $-4.391848\hspace{0.2cm} 10^{-5}$  \\\hline
$b_2$ &$0$                                                       & $+1.564701\hspace{0.2cm} 10^{-7}$  & $+2.764542\hspace{0.2cm} 10^{-7}$  \\\hline
$b_3$ &$0$                                                       & $-2.619560\hspace{0.2cm} 10^{-10}$ & $-6.117961\hspace{0.2cm} 10^{-10}$  \\\hline
$b_4$ &$0$                                                       & $+1.593366\hspace{0.2cm} 10^{-13}$ & $+6.348289\hspace{0.2cm} 10^{-13}$  \\\hline
$b_5$ &$0$                                                      & $0$                                                        & $-3.157181\hspace{0.2cm} 10^{-16}$  \\\hline
$b_6$ &$0$                                                      & $0$                                                        & $+6.143615\hspace{0.2cm} 10^{-20}$  \\\hline
\end{tabular}
\medskip

\newpage

\vspace{-0.3 cm}
\section{Parameters $a_i$ of the $\pi$ cosine distribution in the $\eta$N $\rightarrow$ $\pi$N channel}
\label{Appeta2pi}
\center
$f(cos\theta) = \sum_{i=0}^6 \mathbf{a_i}(P_{Lab})\ cos^i\theta,$\hspace{1cm} with\hspace{1cm}  $\mathbf{a_i}(P_{Lab})=\sum_{i=0}^6 b_i\ P_{Lab}^i$
\medskip

\fbox{$P_{Lab}:$ MeV/c; $\frac{d\sigma}{d\Omega}: mb/sr$} 
\center
\begin{tabular}{|c|c|c|c|c|c|}\hline
$\mathbf{a_0}$    & $P_{Lab} \le 400.$ & $400 < P_{Lab} \le 700.$ & $ 700. < P_{Lab}$  \\\hline\hline
$b_0$ &$+3.830064\hspace{0.2cm} 10^{1}$    & $+6.032010\hspace{0.2cm} 10^{2}$     & $-8.068436\hspace{0.2cm} 10^{1} $  \\\hline
$b_1$ &$-7.469799\hspace{0.2cm} 10^{-1}$   & $-6.737221\hspace{0.2cm}            $    & $+4.653326\hspace{0.2cm} 10^{-1}$  \\\hline
$b_2$ &$+7.230513\hspace{0.2cm} 10^{-3}$  & $+3.123846\hspace{0.2cm} 10^{-2}$   & $-1.093509\hspace{0.2cm} 10^{-3}$  \\\hline
$b_3$ &$-3.862737\hspace{0.2cm} 10^{-5}$   & $-7.669301\hspace{0.2cm} 10^{-5}$   & $+1.354028\hspace{0.2cm} 10^{-6}$  \\\hline
$b_4$ &$+1.155391\hspace{0.2cm} 10^{-7}$  & $+1.049849\hspace{0.2cm} 10^{-7}$   & $-9.341903\hspace{0.2cm} 10^{-10}$ \\\hline
$b_5$ &$-1.813002\hspace{0.2cm} 10^{-10}$ & $-7.593899\hspace{0.2cm} 10^{-11}$ & $+3.408224\hspace{0.2cm} 10^{-13}$ \\\hline
$b_6$ &$+1.160837\hspace{0.2cm} 10^{-13}$& $+2.267918\hspace{0.2cm} 10^{-14}$ & $-5.139366\hspace{0.2cm} 10^{-16}$ \\\hline
\end{tabular}
\smallskip

\begin{tabular}{|c|c|c|c|c|c|}\hline
$\mathbf{a_1}$    & $P_{Lab} \le 500.$ & $500 < P_{Lab} \le 700.$ & $ 700. < P_{Lab}$  \\\hline\hline
$b_0$ &$+2.547230\hspace{0.2cm} 10^{-1}$   & $-5.760562\hspace{0.2cm} $     & $-4.100383\hspace{0.2cm} 10^{1} $  \\\hline
$b_1$ &$+6.516398\hspace{0.2cm} 10^{-4}$   & $+6.894931\hspace{0.2cm} 10^{-2}$   & $+2.203918\hspace{0.2cm} 10^{-1}$  \\\hline
$b_2$ &$-3.564530\hspace{0.2cm} 10^{-6}$    & $-2.480862\hspace{0.2cm} 10^{-4}$   & $-4.845757\hspace{0.2cm} 10^{-4}$  \\\hline
$b_3$ &$+4.934322\hspace{0.2cm} 10^{-8}$   & $+3.599251\hspace{0.2cm} 10^{-7}$   & $+5.644116\hspace{0.2cm} 10^{-7}$  \\\hline
$b_4$ &$-2.342298\hspace{0.2cm} 10^{-10}$  & $-1.824213\hspace{0.2cm} 10^{-10}$ & $-3.686161\hspace{0.2cm} 10^{-10}$ \\\hline
$b_5$ &$+4.113350\hspace{0.2cm} 10^{-13}$ & $0$                                                       & $+1.281122\hspace{0.2cm} 10^{-13}$ \\\hline
$b_6$ &$-2.425827\hspace{0.2cm} 10^{-16}$ & $0$                                                       & $-1.851188\hspace{0.2cm} 10^{-17}$ \\\hline
\multicolumn{3}{c}{}\\ 
\end{tabular}
\begin{tabular}{|c|c|c|c|c|c|}\hline
$\mathbf{a_2}$    & $P_{Lab} \le 550.$ & $550 < P_{Lab} \le 700.$ & $ 700. < P_{Lab}$  \\\hline\hline
$b_0$ &$+1.524349\hspace{0.2cm} 10^{-1}$   & $-5.116601\hspace{0.2cm} $              & $+8.084776\hspace{0.2cm} 10^{1} $  \\\hline
$b_1$ &$-4.745692\hspace{0.2cm} 10^{-3}$    & $+4.108704\hspace{0.2cm} 10^{-2}$ & $-4.775194\hspace{0.2cm} 10^{-1}$  \\\hline
$b_2$ &$+6.996373\hspace{0.2cm} 10^{-5}$   & $-8.734112\hspace{0.2cm} 10^{-5}$  & $+1.146234\hspace{0.2cm} 10^{-3}$  \\\hline
$b_3$ &$-2.759605\hspace{0.2cm} 10^{-7}$    & $+5.514651\hspace{0.2cm} 10^{-8}$ & $-1.441294\hspace{0.2cm} 10^{-6}$  \\\hline
$b_4$ &$+4.624668\hspace{0.2cm} 10^{-10}$ & $0$                                                      & $+1.005796\hspace{0.2cm} 10^{-9}$ \\\hline
$b_5$ &$-3.030435\hspace{0.2cm} 10^{-13}$  & $0$                                                      & $-3.701960\hspace{0.2cm} 10^{-13}$ \\\hline
$b_6$ &$+1.352952\hspace{0.2cm} 10^{-17}$ & $0$                                                      & $+5.621795\hspace{0.2cm} 10^{-17}$ \\\hline
\multicolumn{3}{c}{}\\
\end{tabular}
\begin{tabular}{|c|c|c||c|c|c|c|}\hline
$\mathbf{a_3}$  & $P_{Lab} \le 700.$                               & $700 < P_{Lab} $                                & $\mathbf{a_4}$& $P_{Lab} \le 550.$                                 & $550 < P_{Lab} \le 650.$                 & $650 < P_{Lab}$   \\\hline\hline
$b_0$                & $+1.270435\hspace{0.2cm} 10^{-1}$ & $+1.552846\hspace{0.2cm} 10^{2}$  &$b_0$                 &$-5.633076\hspace{0.2cm} 10^{-2} $ & $-4.482122\hspace{0.2cm} $            & $-2.447717\hspace{0.2cm}$                    \\\hline
$b_1$                &$-4.735559\hspace{0.2cm} 10^{-3}$   & $-9.323442\hspace{0.2cm} 10^{-1}$ &$b_1$                &$+2.109593\hspace{0.2cm} 10^{-3}$  & $+1.827203\hspace{0.2cm} 10^{-2}$  &$+6.530743\hspace{0.2cm} 10^{-2}$    \\\hline
$b_2$                &$+5.903545\hspace{0.2cm} 10^{-5}$  & $+2.261028\hspace{0.2cm} 10^{-3}$ &$b_2$                &$-2.631251\hspace{0.2cm} 10^{-5}$  & $-1.698136\hspace{0.2cm} 10^{-5}$ &$-2.621981\hspace{0.2cm} 10^{-4}$    \\\hline
$b_3$                &$-3.407333\hspace{0.2cm} 10^{-7}$   & $-2.867416\hspace{0.2cm} 10^{-6}$ &$b_3$                &$+1.353545\hspace{0.2cm} 10^{-7}$  &$0$                           
                                                           & $+4.452787\hspace{0.2cm} 10^{-7}$    \\\hline
$b_4$                &$+9.783322\hspace{0.2cm} 10^{-10}$& $+2.015156\hspace{0.2cm} 10^{-9}$ &$b_4$                &$-3.166229\hspace{0.2cm} 10^{-10}$ &$0$                      
                                                           & $-3.820460\hspace{0.2cm} 10^{-10}$    \\\hline
$b_5$                &$-1.356389\hspace{0.2cm} 10^{-12}$ & $-7.459580\hspace{0.2cm} 10^{-13}$&$b_5$               &$+3.858551\hspace{0.2cm} 10^{-13}$ &$0$                       
                                                          & $+1.640033\hspace{0.2cm} 10^{-13}$    \\\hline
$b_6$                &$+7.061866\hspace{0.2cm} 10^{-16}$ &$+1.138088\hspace{0.2cm} 10^{-16}$&$b_6$               &$-2.051840\hspace{0.2cm} 10^{-16}$ &$0$                       
                                                          & $-2.808337\hspace{0.2cm} 10^{-17}$    \\\hline
\multicolumn{3}{c}{}\\ 
\end{tabular}
\begin{tabular}{|c|c|c||c|c|c|}\hline
$\mathbf{a_5}$  & $P_{Lab} \le 700.$                               & $700 < P_{Lab} $                                  & $\mathbf{a_6}$& $P_{Lab} \le 600.$                              & $600 < P_{Lab}$   \\\hline\hline
$b_0$                & $-6.810842\hspace{0.2cm} 10^{-2}$ & $-4.328028\hspace{0.2cm} 10^{1}$     &$b_0$               & $+2.418893\hspace{0.2cm} 10^{-3}$ & $+1.426952\hspace{0.2cm} $                   \\\hline
$b_1$                &$+2.552380\hspace{0.2cm} 10^{-3}$ & $+2.757524\hspace{0.2cm} 10^{-1}$  &$b_1$                &$-6.081534\hspace{0.2cm} 10^{-5}$   & $-9.167580\hspace{0.2cm} 10^{-3}$        \\\hline
$b_2$                &$-3.234292\hspace{0.2cm} 10^{-5}$  & $-6.953576\hspace{0.2cm} 10^{-4}$  &$b_2$                &$+5.955500\hspace{0.2cm} 10^{-7}$  & $+2.385312\hspace{0.2cm} 10^{-5}$       \\\hline
$b_3$                &$+1.865842\hspace{0.2cm} 10^{-7}$ & $+9.065678\hspace{0.2cm} 10^{-7}$   &$b_3$               &$-2.947343\hspace{0.2cm} 10^{-9}$   & $-3.237490\hspace{0.2cm} 10^{-8}$        \\\hline
$b_4$                &$-5.344420\hspace{0.2cm} 10^{-10}$& $-6.503137\hspace{0.2cm} 10^{-10}$ &$b_4$               &$+7.812226\hspace{0.2cm} 10^{-12}$ & $+2.428560\hspace{0.2cm} 10^{-11}$      \\\hline
$b_5$                &$+7.397533\hspace{0.2cm} 10^{-13}$& $+2.445059\hspace{0.2cm} 10^{-13}$ &$b_5$              &$-1.063594\hspace{0.2cm} 10^{-14}$ & $-9.570613\hspace{0.2cm} 10^{-15}$      \\\hline
$b_6$                &$-3.858406\hspace{0.2cm} 10^{-16}$& $-3.775268\hspace{0.2cm} 10^{-17}$  &$b_6$              &$+5.721872\hspace{0.2cm} 10^{-18}$ & $+1.549323\hspace{0.2cm} 10^{-18}$   \\\hline
\end{tabular}
\medskip

\newpage

%


\begin{thebibliography}{00}
%
%

\bibitem{BLA93}
M. Blann {\it et al.}, International Code Comparison for Intermediate Energy Nuclear Data. NEA/OECD, NSC/DOC(94)-2, Paris, 1993.

\bibitem{MIC97}
R. Michel et P. Nagel. International Codes and Model Intercomparison for Intermediate Energy Activation Yields. NEA/OECD, NSC/DOC(97)-1, Paris, 1997.

\bibitem{LER11}
S. Leray {\it et al.}, J. Korean Phys. Soc. {\bf 59}, 791, doi: 10.3938/jkps.59.791, (2011).

\bibitem{NIL87}
B. Nilsson-Almqvist and E. Stenlund, Computer Physics Commun. {\bf 43}, 387 (1987).

\bibitem{SJO06}
T. Sj\"{o}strand, S. Mrenna and P. Skands,  arXiv:hep-ph/0603175v2, March 2006

\bibitem{AIC85}
J. Aichelin and C. M. Ko, Phys. Rev. Lett. {\bf 55}, 2661 (1985).

\bibitem{LI89}
Q. Li, J. Q. Wu, and C. M. Ko, Phys. Rev. {\bf C 39}, 849 (1989).

\bibitem{BUS12}
O. Buss {\it et al.}, Phys. Rep. {\bf 512}, 1–124 (2012).

\bibitem{AIC91}
J. Aichelin, Phys. Rep. (Review Section of Physics Letters)  {\bf 202}, 233-360 (1991).
 
\bibitem{NII95}
K. Niita {\it et al.}, Phys. Rev. {\bf C 52}, 2620-2635 (1995).
 
\bibitem{BAS98}
S. A. Bass {\it et al.}, Prog. Part. Nucl. Phys. {\bf 41}, 225-370 (1998).

\bibitem{HAR98}
Ch. Hartnack {\it et al.}, Eur. Phys. J.  {\bf A 1}, 151-169 (1998).

\bibitem{DAV15}
J.-C. David, Eur. Phys. J.  {\bf A 51},  68 (2015).

\bibitem{PED11}
S. Pedoux and J. Cugnon, Nucl. Phys. {\bf A 866}, 16-36 (2011).

\bibitem{BOU13}
A. Boudard  {\it et al.}, Phys. Rev. {\bf C 87}, 014606 (2013).

\bibitem{MAN14}
D. Mancusi {\it et al.}, Phys. Rev. {\bf C 90}, 054602 (2014).

\bibitem{MAN17}
D. Mancusi {\it et al.}, Eur. Phys. J. {\bf A 53}, 80 (2017).

\bibitem{KUP11}
A. Kupsc and A. Wirzba, Journal of Physics: Conference Series {\bf 335}, 012017 (2011).

\bibitem{GAT16}
C. Gatto, B. Fabela Enriquez and M. I. Pedraza Morales, Proceedings of Science (ICHEP2016), 812 (2016). 


\bibitem{RIC70}
W. B. Richards {\it et al.}, Phys. Rev. {\bf D 1}, 10 (1970).

\bibitem{PRA05}
S. Prakhov {\it et al.}, Phys. Rev. {\bf C 72}, 015203 (2005).

\bibitem{BRO79}
R. M. Brown {\it et al.}, Nucl. Phys. {\bf B153}, 89 (1979).

\bibitem{DEI69}
W. Deinet {\it et al.}, Nucl. Phys. {\bf B11}, 495 (1969).

\bibitem{DUR08}
J. Durand {\it et al.}, Phys. Rev. {\bf C 78}, 025204 (2008).

\bibitem{CUG90}
J. Cugnon {\it et al.}, Phys. Rev. {\bf C 41}, 1701-1718 (1990).

\bibitem{BAL88}
A. Baldini  {\it et al.} “Total Cross Sections for Reactions of High Energy Particles”, Landolt-B\"{o}rnstein Numerical Data and Functional Relationships in Science and Technology. (Springer- Verlag, Berlin, 1988), Vol. 12.

\bibitem{CAL98}
H. Cal\'en {\it et al.}, Phys. Rev. {\bf C 58}, 2667-2670 (1998).

\bibitem{HAN04}
C. Hanhart {\it et al.}, Phys. Rep. {\bf 397}, 155-256 (2004).

\bibitem{SIB97}
A. Sibirtsev {\it et al.}, Z. Phys. {\bf A 358}, 357-367 (1997).

\bibitem{SMY00}
J. Smyrski {\it et al.}, Phys. Lett. {\bf B474}, 182-187 (2000).

\bibitem{CAL96}
H. Cal\' en {\it et al.}, Phys. Lett. {\bf B366}, 39-43 (1996).

\bibitem{BER93}
A. M. Bergdolt {\it et al.}, Phys. Rev. {\bf D 48}, R2969 (1993).

\bibitem{HIB98}
F. Hibou {\it et al.}, Phys. Lett. {\bf B438}, 41-46 (1998).

\bibitem{CHI94}
E. Chiavassa {\it et al.}, Phys. Lett. {\bf B322}, 270-274 (1994).

\bibitem{CAL98}
H. Cal\' en {\it et al.}, Phys. Rev.  {\bf C 58}, 2667 (1998).

\bibitem{CAL97}
H. Cal\' en {\it et al.}, Phys. Rev. Lett. {\bf 79}, 2642 (1997).

\bibitem{CAL98b}
H. Cal\' en {\it et al.}, Phys. Rev. Lett. {\bf 80}, 2069 (1998).

\bibitem{SIB96}
A. Sibirtsev {\it et al.}, Nucl. Phys. {\bf A 604}, 455-465 (1996).

\bibitem{BAR04}
S. Barsov {\it et al.}, Eur. Phys. J. {\bf A 21}, 521-527 (2004).

\bibitem{HIB99}
F. Hibou {\it et al.}, Phys. Rev. Lett. {\bf 83}, 492 (1999).

\bibitem{ABD01}
S. Abd-El-Samad {\it et al.}, Phys. Lett. {\bf B522}, 16-21 (2001).

\bibitem{RUS10}
A. Rustamov {\it et al.},"Inclusive meson production in 3.5 GeV pp collisions studied with the HADES spectrometer", AIP Conference Proceedings, 1257, 736-740 (2010), DOI:http://dx.doi.org/10.1063/1.3483432

\bibitem{KAM13}
H. Kamano {\it et al.}, Phys. Rev.  {\bf C 88}, 035209 (2013).

\bibitem{DUD76}
G. N. Dudkin {\it et al.}, JETP Letters  {\bf 23}, 77 (1976).

\bibitem{LYK99}
G. I. Lykasov {\it et al.}, Eur. Phys. J. {\bf A 6}, 71-81 (1999).

\bibitem{DEB75}
N. C. Debenham {\it et al.}, Phys. Rev. {\bf D 12}, 2545 (1975).

\bibitem{VET91}
T. Vetter {\it et al.}, Phys. Lett. {\bf B263}, 153-156 (1991).

\bibitem{CUG97}
J. Cugnon {\it et al.}, Phys. Rev. {\bf C 56}, 2431-2439 (1997).

\bibitem{PDG14}
Particle Data Group, Chinese Physics {\bf C 38}, 090001 (2014).

\bibitem{ZHO06}
X. H. Zhong {\it et al.}, Phys. Rev. {\bf C 73}, 015205 (2006).

\bibitem{NAG05}
H. Nagahiro {\it et al.}, Nucl. Phys. {\bf A 761}, 92-119 (2005).

\bibitem{FRI14}
S. Friedrich {\it et al.}, Phys. Lett. {\bf B736}, 26-32 (2014).

\bibitem{MET17}
V. Metag {\it et al.}, EPJ Web of Conferences {\bf 134}, 03003 (2017).

\bibitem{CAT08}
M.G. Catanesi {\it et al.}, Phys. Rev. {\bf C 77}, 055207 (2008).

\bibitem{GOL93}
Ye. S. Golubeva {\it et al.}, Nucl. Phys. {\bf A 562}, 389 (1993).

\bibitem{AGA13}
G. Agakishiev {\it et al.}, Phys. Rev. {\bf C 88}, 024904 (2013).

\end{thebibliography}
\end{document}